\documentclass[aps,pra,reprint,showpacs,showkeys,groupedaddress,nofootinbib,superscriptaddress]{revtex4-1}
\usepackage{hyperref}
\usepackage[T1]{fontenc}
\usepackage[ansinew]{inputenc}
\usepackage{amsmath}
\usepackage{setspace}
\usepackage{psfrag}
\usepackage{mathtools}
\usepackage{amssymb,lineno,amsfonts}
\usepackage{verbatim}
\usepackage{color}
\usepackage{textcomp}
\usepackage[toc,page]{appendix}
\usepackage{graphicx}
\usepackage[caption=false]{subfig}
\usepackage{float}
\usepackage{bm}
\usepackage{dcolumn}
\usepackage{gensymb}
\usepackage{array}
\usepackage{ifpdf}
\usepackage{bbm}
\usepackage{mathrsfs}
\usepackage{upgreek}
\usepackage{epstopdf}
\usepackage{esvect}
\usepackage[usenames,dvipsnames]{xcolor}
\definecolor{med-blue}{RGB}{25,25,112}
\hypersetup{colorlinks, linkcolor={blue},citecolor={blue}, urlcolor={blue}}
\usepackage[english]{babel}
\usepackage[autostyle, english = american]{csquotes}
\usepackage{enumerate}
\usepackage{soul}
\usepackage{amsfonts}
\usepackage{amssymb}
\begin{document}
\title{A general variational approach for equilibrium phase boundaries of trapped spin-1 Bose-Einstein condensates}

\author{Sahil Satapathy}
\email{sahil.satapathy@students.iiserpune.ac.in}
\affiliation{Department of Physics, Indian Institute of Science Education and Research, Dr Homi Bhabha Road, Pune 411 008, India}

\author{Projjwal K. Kanjilal}
\email{projjwal.kanjilal@students.iiserpune.ac.in}
\affiliation{Department of Physics, Indian Institute of Science Education and Research, Dr Homi Bhabha Road, Pune 411 008, India}
\affiliation{Tata Institute of Fundamental Research, Tata Institute of Fundamental Research, Hyderabad, 500 046, India}

\author{A. Bhattacharyay}
\email{a.bhattacharyay@iiserpune.ac.in}
\affiliation{Department of Physics, Indian Institute of Science Education and Research, Dr Homi Bhabha Road, Pune 411 008, India}

\date{\today}

\begin{abstract}{We develop a simple and general variational method to estimate the solutions of the Gross-Pitaevskii equations and obtain the corresponding density profiles for all spin states of a trapped spin-1 Bose-Einstein condensate. We further employ this approach to obtain the complete phase diagram of the system under quasi-one-dimensional harmonic confinement, with ferromagnetic or antiferromagnetic spin interactions. We identify a suitable scaling that collapses all phase diagrams for different system sizes (i.e., total particle number) into a universal (system size-independent) phase diagram. The complete phase diagram for a confined system shows some significant qualitative differences compared to that of a condensate with homogeneous density distribution. The phase diagrams reported here could help identify the important parameter regimes in which phase transitions in the confined system, in general, occur. This knowledge of the region of phase boundaries can enable a reliable investigation of the instabilities near the boundaries that drive phase transitions.}




\end{abstract}

\maketitle

\section{Introduction}
The experimental observation of the Bose-Einstein condensate (BEC) \cite{1stexps1,1stexps2,1stexps3} nearly thirty years ago marked a paradigm shift in atomic physics. One can explore the concepts from condensed matter physics with unprecedented control in ultracold quantum gases. The subsequent arrival of the optical trapping technique enabled confinement of atoms with all internal spin states, giving rise to much richer spinor-BEC \cite{spinorexp1,spinorexp2}. Spinor-BEC emerged as a key platform that realises unique spin-domain forming structures \cite{stenger98,kibble-zurek1,PhysRevA.78.023632,PhysRevA.60.4857,PhysRevA.85.023601,jimenez2018spontaneous,Kanjilal_2020,*Kanjilal_corri,PhysRevResearch.3.023043}, exotic spin-textures \cite{spin-textures1,spin_textures_exp,spin_textures_muller,Ueda_2014,PhysRevLett.103.250401,PhysRevA.102.023326,Choi_2012, PhysRevA.105.053303, Mistakidis_nature} and nonlinear excitations such as multi-component solitons in spinor condensates \cite{cpl_soliton,cpl_soliton2,solitontheo1,solitontheo2}.

In a spinor-BEC, the interplay of spin-dependent and spin-independent interactions, along with the externally tunable linear and quadratic Zeeman fields, leads to a rich phase diagram \cite{T-L-Ho,KAWAGUCHI2012253,RevModPhys.85.1191} featuring multiple quantum phase transitions. Understanding the nature of instabilities and their interplay near the phase boundaries of a trapped spin-1 Bose-Einstein condensate requires, first and foremost, determining the location of these phase boundaries in the parameter space. However, a general method of locating the phase boundaries in the relevant parameter space for a trapped condensate does not yet exist, like the one that exists for a homogeneously extended (uniform) condensate. The primary difficulty comes from the lack of exact solvability of the Gross-Pitaevskii equation, which is formulated as a nonlinear Schr\"{o}dinger equation. Various attempts to locate the region of phase boundaries between spin states exist in the literature \cite{stenger98,mats10gr,Zhang_2003,PhysRevA.85.023601}, although they are based on specific situations.

Phase boundaries are regions where instabilities play a major role, defining the ensuing phase transitions. A precise knowledge of their location, therefore, forms the foundation for understanding the dynamics of fluctuations in the system's free energy. In the context of a spinor-BEC, one generally knows phase boundaries in the space of the linear ($p$) and quadratic ($q$) Zeeman parameter for a homogeneous condensate under various broad conditions of spin-spin interactions (for example, ferro and anti-ferro magnetic) \cite{T-L-Ho,KAWAGUCHI2012253}. However, these results have limited applicability to experimentally relevant trapped systems with a finite number of particles. In the presence of confinement, the phase boundaries in the $p$-$q$ plane are not only often significantly shifted but can also differ qualitatively from their homogeneous counterparts. This naturally exhibits the natural interplay between particle number, trap geometry, and interaction strength. It is worth noting that the chemical potential also changes significantly for a trapped condensate in relation to that of a homogeneous one \cite{2003bose,Pethick_Smith_2008}.

Although the Thomas-Fermi approximation (TFA) \cite{2003bose} and single-mode approximation (SMA) \cite{SMA1,SMA2,SMA_physics,validity_SMA} are widely used methods for obtaining the condensate profile under trapping, they have severe limitations. For example, in states with more than one component, i.e., multi-component stationary states, the TFA is inaccurate at predicting domain-forming structures. On top of that TFA, by construction, is better suited for condensates with a large number of particles. This limits the estimation of the condensate profile for a small number of particles using TFA. 

Similarly, using SMA can lead to overestimation of some components and underestimation of others when dealing with multi-component states. These limitations have already been pointed out in \cite{kanjilal_variational, Kanjilal_PRA}, where a variational method was proposed by Kanjilal and Bhattacharyay that works for different condensate sizes under harmonic confinement. The method was shown to produce number densities in excellent agreement with the numerical simulations under specific constraints. Although that method proved useful, it became difficult to obtain an accurate estimate in regions where the linear and quadratic Zeeman terms became very small. 

The method \cite{kanjilal_variational,Kanjilal_PRA} could become somewhat cumbersome, especially when estimating the free energy difference of phases for small $p$, $q$ to achieve a smooth phase boundary. For example, the phase boundary between polar and antiferromagnetic states for the antiferromagnetic type of spin-spin interaction could not be estimated with high accuracy using the previous method. In pursuit of a general and relatively simple variational method to demarcate the regions of phase transitions for a trapped spinor BEC, we have now developed a new variational scheme that differs significantly from previous ones and also enables us to reduce implementation complexity. 

\par
Let us highlight some of the significant results obtained with the new, more general variational method we report. We determine the phase boundaries between all stationary states in the $p$-$q$ parameter space by varying the total number of condensate particles and the spin interaction strength under a quasi-one-dimensional harmonic confinement. We identify the scaling of the linear and quadratic Zeeman terms with particle number, collapsing the phase boundaries into universal curves for both ferromagnetic and antiferromagnetic spin interactions. We observe that, irrespective of the type of spin-interaction, this scaling depends as $~N^{2/3}$, where $N$ is the number of particles. We identify that this scaling occurs naturally in the quasi-one-dimensional confined system due to transverse confinement.

For antiferromagnetic interactions, the phase boundary of the trapped condensate differs significantly from the homogeneous case: the phase boundary separating the ferromagnetic and antiferromagnetic states for a positive $q$ exhibits distinct behaviour under confinement. For the other regime of ferromagnetic type of spin interaction, the phase-matched state, characterised by a non-zero occupation of all the spin components, appears only for a very narrow region of positive $q$. This is in sharp contrast to the homogeneous case, where the phase-matched state is energetically favoured even for large $q$. As a consequence, in a trapped system, tuning the quadratic Zeeman field $q$ can introduce a direct transition from the polar to the ferromagnetic phase, whereas in a homogeneous system, such a transition is forbidden and must proceed through the phase-matched state. Note that the variational method introduced in this article is fairly general and can be extended to higher spin systems.

The paper is organised as follows: In Section II, we discuss the mean-field theory leading to the homogeneous phase diagram. In Section III, we discuss the variational method introduced in this article. This method is also used in Section IV to obtain a complete phase diagram in the presence of external confinement when the spin interaction is both positive and negative.

\section{Mean-field theory}

\begin{figure*}[htb]
  \centering
  \subfloat[]{
    \includegraphics[width=0.32\textwidth]{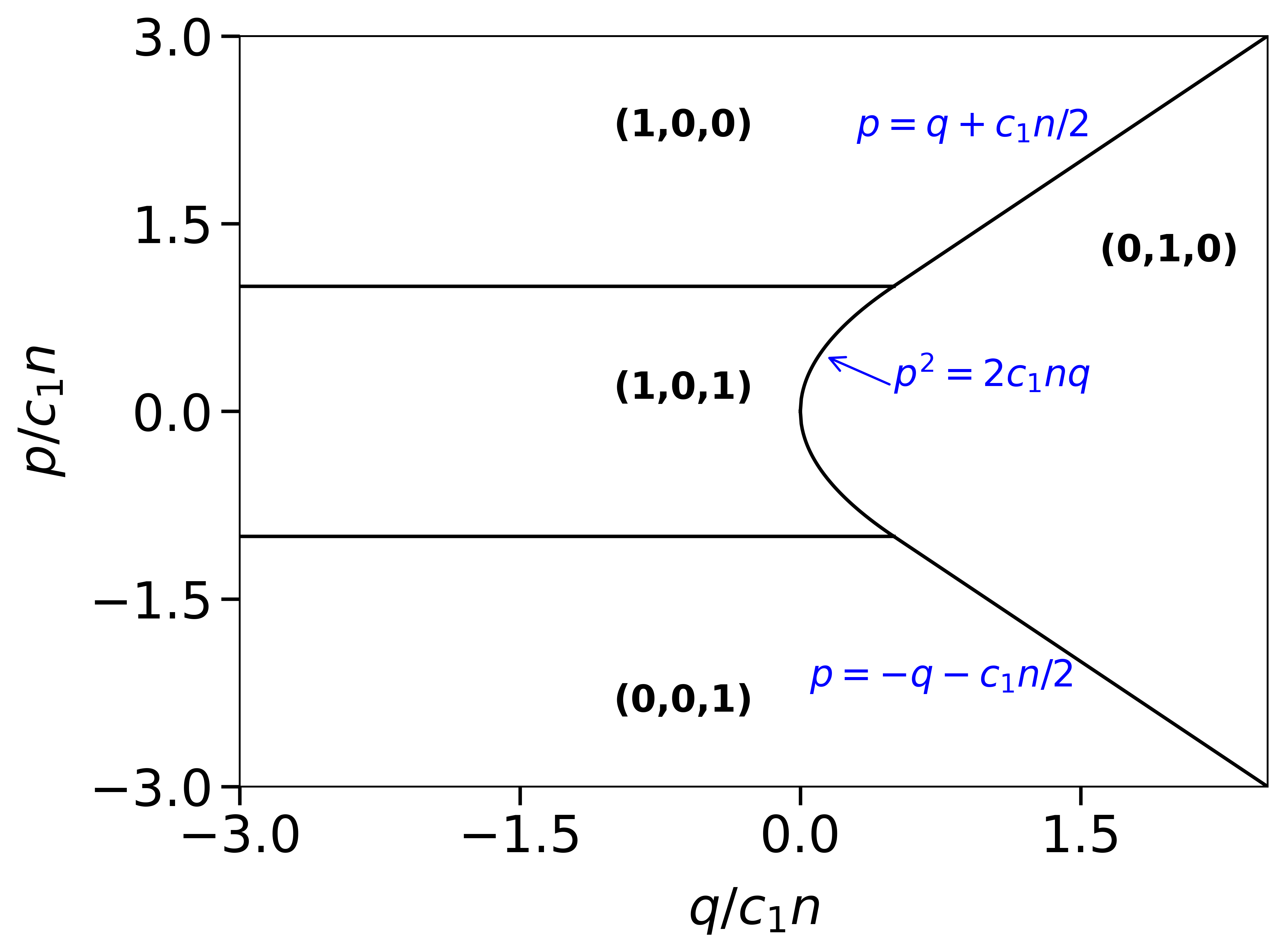}
    \label{fig:1a}
  }
  \subfloat[]{
    \includegraphics[width=0.32\textwidth]{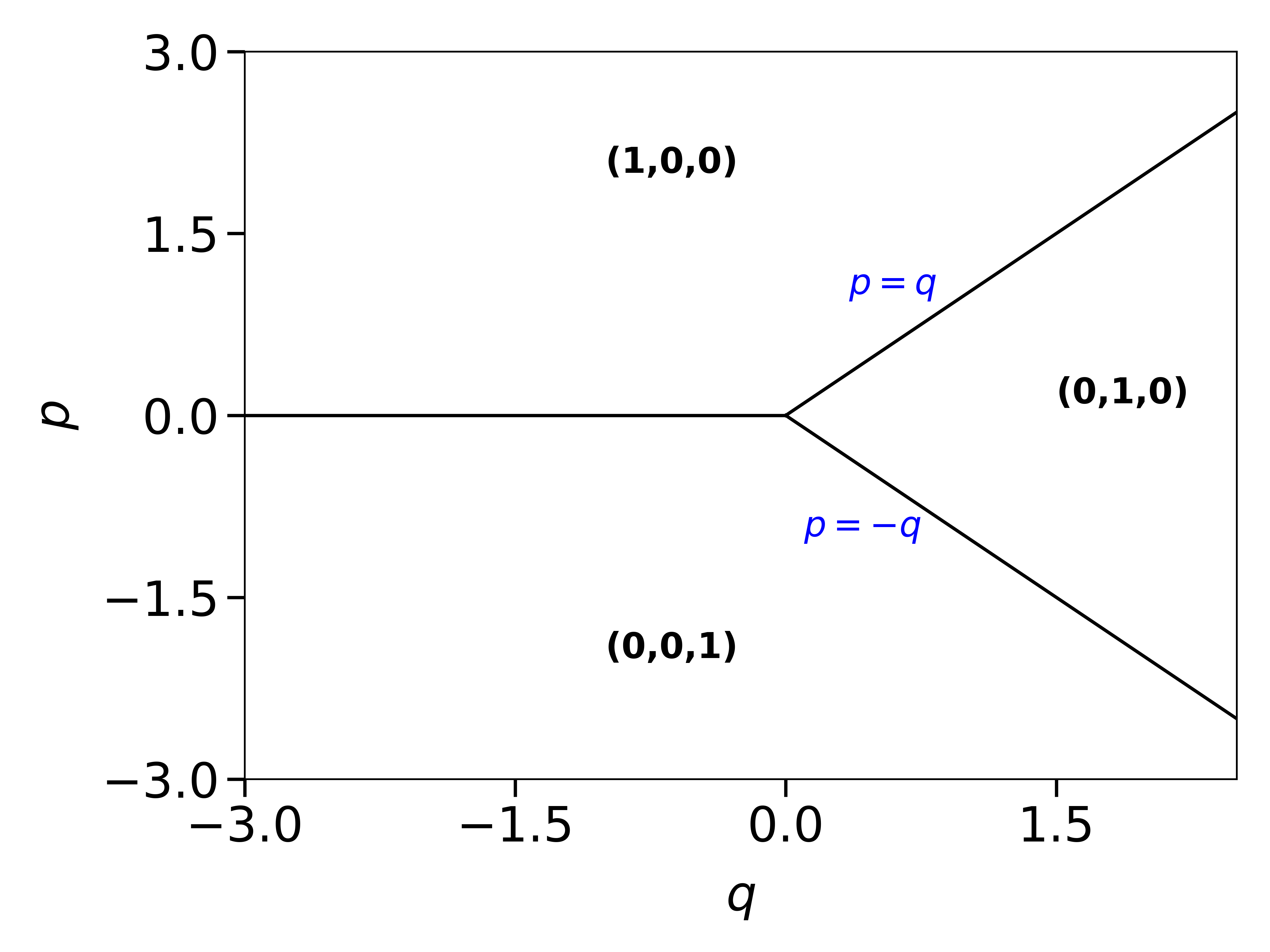}
    \label{fig:1b}
  }
  \subfloat[]{
    \includegraphics[width=0.32\textwidth]{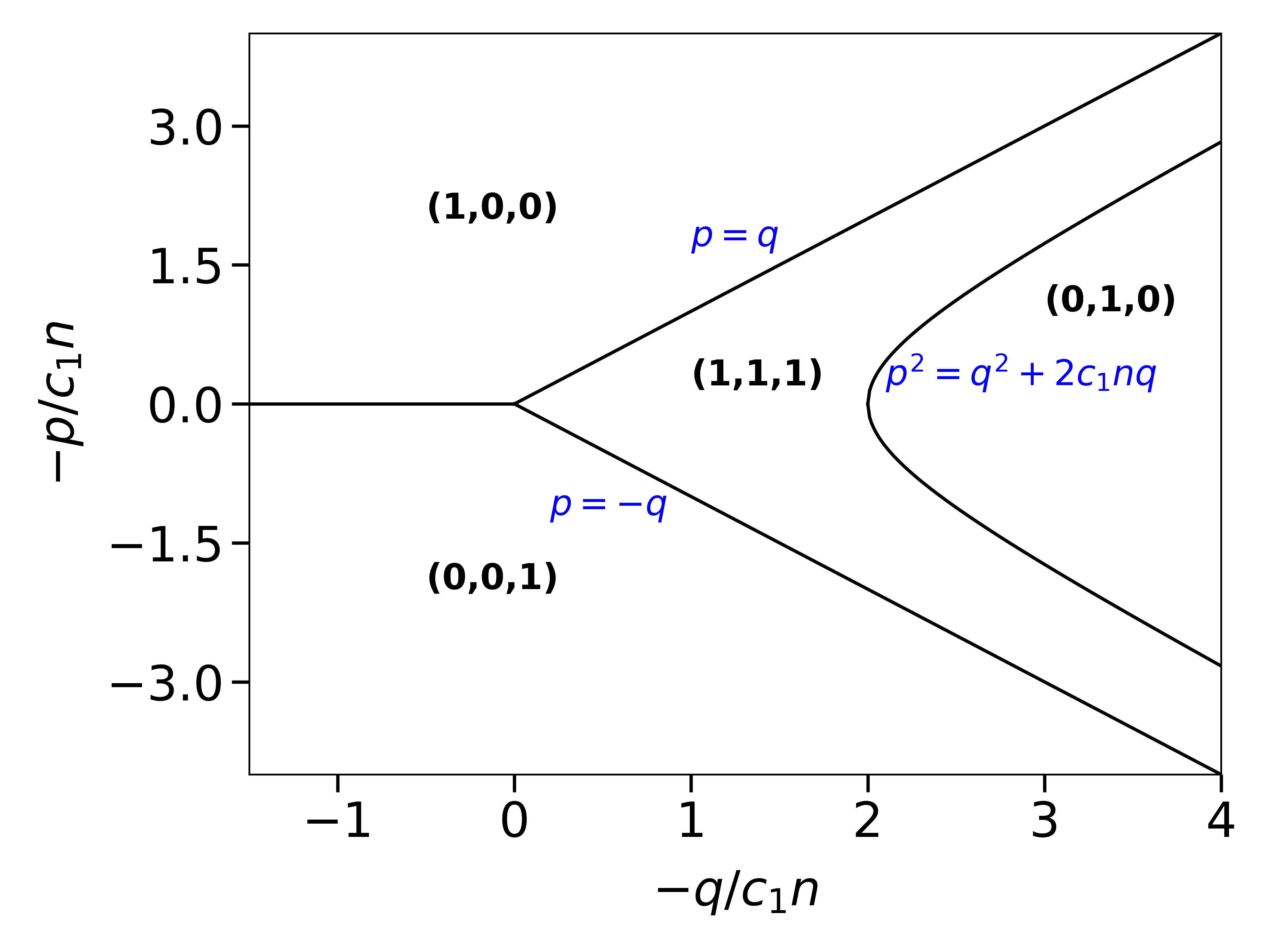}
    \label{fig:1c}
  }
  \caption{Phase diagram in linear ($p$) and quadratic ($q$) Zeeman terms for spin-1 condensate in the absence of any trapping potential, for (a) antiferromagnetic-type, (b) in-the-absence-of, and (c) ferromagnetic-type of spin interaction.}
  \label{fig:1}
\end{figure*}
In the presence of the contact-interaction, the spin-1 BEC is well-described by the mean-field Hamiltonian,
\begin{equation}
\begin{split}
H = \int d\bold{r}\sum_{m=-1}^1\Bigg[ \psi_m^*\Big(-\dfrac{\hbar^2}{2M}\bold{\nabla}^2&+U(\bold{r})-pm+qm^2\Big)\psi_m\\
&+ \dfrac{c_0}{2}n^2 + \dfrac{c_1}{2} |\bold{F}|^2\Bigg],
\end{split}
\end{equation}
where the the mean-fields $\psi_m$ for spin-projections $m = 0, 1, -1$, in general depends on space and time, i.e., $\psi_m(\bold{r},t)$. The term $U(\bold{r})$ represents the external trapping potential and $n$ is the total density. The parameters $p$ and $q$ represent the linear and quadratic Zeeman fields, respectively, that couple to the system's spin projection. The contact interaction can be decomposed into the density-density interaction with strength $c_0$ and a spin-dependent interaction with strength $c_1$. The operator $\bold{F}$ is defined using the spin-1 matrices $f^{x,y,z}_{mm'}$ \cite{KAWAGUCHI2012253} as,
\begin{equation}\label{eq2}
    \bold{F}=\sum^{1}_{m,m'=-1} \psi^{*}_m(\bold{r},t) \bold{f}_{mm'}\psi_{m'}(\bold{r},t).
\end{equation}
The dynamics of the mean-fields $\psi_m(\bold{r},t)$ of the spin-1 BEC is well described by the Gross-Pitaevskii equations (GPE)  \cite{doi:10.1143/JPSJ.67.1822,T-L-Ho,KAWAGUCHI2012253},
\begin{equation}\label{eq1}
    \begin{split}
        i\hbar \dfrac{\partial\psi_m}{\partial t} = \Big(-\dfrac{\hbar^2}{2M}\bold{\nabla}^2&+U(\bold{r})-pm+qm^2 +c_0n\Big)\psi_m\\
        &+ c_1 \sum_{m'=-1}^1 \bold{F}\cdot \bold{f}_{mm'} \psi_{m'},
    \end{split}
\end{equation}
which is obtained from the mean-field Hamiltonian.
We use the ansatz,
\begin{equation}\label{eq3}
    \psi_m(\bold{r},t)=\sqrt{n_m(\bold{r},t)}exp\big(-i\theta_m(t)\big)exp\Big(-\dfrac{i\mu t}{\hbar}\Big)    
\end{equation}
representing the mean-fields in terms of the sub-component densities $n_m(\bold{r},t)$, phases $\theta_m$ and the chemical potential $\mu$, to reduce the GPEs to,
\begin{equation}\label{eq4}
    \dot{n}_0=-\dfrac{4c_1 n_0 \sqrt{n_1n_{-1}}\sin\theta_r}{\hbar},
 \end{equation}
 \begin{equation}\label{eq5}
    \dot{n}_{\pm1}=\dfrac{2c_1 n_0 \sqrt{n_1n_{-1}}\sin\theta_r}{\hbar},
 \end{equation}
 \begin{equation}\label{eq6}
    \begin{split}
        \hbar\dot{\theta}_0=\dfrac{1}{\sqrt{n_0}}&\left(\mathcal{H}-\mu\right)\sqrt{n_0}\\
        &+c_1\left(n_1+n_{-1}+2 \sqrt{n_1n_{-1}}\cos\theta_r\right),
    \end{split}
 \end{equation}
 \begin{equation}\label{eq7}
    \begin{split}
        \hbar\dot{\theta}_{\pm1}=\dfrac{1}{\sqrt{n_{\pm1}}}&\left(\mathcal{H}-\mu\right)\sqrt{n_{\pm1}}\pm c_1\left(n_1-n_{-1}\right)+q\\
        &\quad \mp p+c_1n_0\left(1 +\sqrt{\dfrac{n_{\mp1}}{n_{\pm1}}}\cos\theta_r\right),
    \end{split}
 \end{equation}
  where we have suppressed the spatial and temporal dependence of the densities and phases for the notational compactness and defined $\mathcal{H}=-\frac{\hbar^2}{2M}\bold{\nabla}^2 + U(\bold{r})+c_0 n(\bold{r},t)$ and the relative phase $\theta_r\equiv\theta_1+\theta_{-1}-2\theta_0$. We assume that the sub-components are slowly varying so that the second derivative can be neglected.
  \par

  For a stationary state, the sub-component densities and the phases do not have any time dependence and the left-hand side of Eqs.\ref{eq4}-\ref{eq7} vanishes. One can solve the GPEs to obtain the structure of the stationary states and the corresponding density profiles under the Thomas-Fermi approximation, as detailed in \cite{Kanjilal_2020,*Kanjilal_corri}.

 \par
  For the simplest case of homogeneous condensate (in the absence of external trapping $U(\bold{r})$), one can get to the phase diagram of the system in an independently-tunable parameter space of $p$ and $q$ \cite{KAWAGUCHI2012253} as shown in Fig.\ref{fig:1}. We denote the possible stationary states as $(\mathbbm{n}_1,\mathbbm{n}_0,\mathbbm{n}_{-1})$, where if the $m^{th}$ sub-component is populated $\mathbbm{n}_m =1$, and $0$ if it is empty. So, for example, $(0,1,0)$ would be the polar state where only the $m=0$ sub-component is populated, and others are empty. For the anti-ferromagnetic type of spin interaction, i.e., $c_1>0$, one can have a rich phase diagram showing many possible states like ferromagnetic states ($(1,0,0)$ and $(0,0,1)$), anti-ferromagnetic state, i.e., $(1,0,1)$, and polar state existing as ground state for some preferred parameter region of $p$ and $q$ as shown in Fig.\ref{fig:1}\subref{fig:1a}. For the ferromagnetic coupling i.e., $c_1<0$, a unique phase $(1,1,1)$ where all the sub-components are populated with relative phase $\theta_r=0$, hence the name phase-matched (PM) state, emerge as the ground state in some parameter regime with $q>0$, apart from the ferromagnetic and polar state (Fig.\ref{fig:1}\subref{fig:1c}). Even in the absence of a spin-interaction, the Zeeman terms play a crucial role in choosing the ground state (Fig. \ref{fig:1}\subref{fig:1b}). This shows that, even in the absence of trapping, the spin-1 system exhibits a rich phase diagram of magnetic orderings and associated quantum phase transitions (at $T=0$). 
\par 
    In practice, experiments are done with a BEC trapped by a potential. This causes the phase diagrams to change and become richer than the homogeneous case. To reach such a phase diagram, one needs to accurately estimate the phase boundaries, which also helps in understanding the nature of quantum phase transitions. With this goal in mind, we develop a variational method that enables us to describe the trapped phases.


\section{Variational method}
In this work, we will utilise harmonic trapping in a quasi-one-dimensional geometry. We do so to compare our results with numerical simulations, which are easy to implement and reliable in this geometry. Based on these validations of the variational results through numerical simulations, we intend to extend the variational method to higher-dimensional systems in future work. We assume the system to be elongated along the $z$-axis. This would happen if the harmonic trapping frequencies in the other two directions are much greater than that along the $z$-axis, i.e., $\omega^2_z << \omega_x \omega_y$. This allows for integrating out the degrees of freedom in the tighter directions, thereby renormalising some parameters. 

We represent the parameters in the following dimensionless form:
\begin{equation}\label{eq8}
    c_0=2 \pi l^2_{xy} l_z\lambda_0\hbar\omega_z, \quad c_1=2 \pi l^2_{xy} l_z\lambda_1\hbar\omega_z,
\end{equation}
\begin{equation}\label{eq9}
    u_m=2 \pi l^2_{xy} l_z n_m, \quad r=l_z\zeta
\end{equation}

where, $l_z^2=\hbar/(m\omega_z)$, $l_{xy}^2=\hbar/(m\sqrt{\omega_{x}\omega_{y}})$. This makes $\lambda_0$ and $\lambda_1$ a dimensionless analog of $c_0$ and $c_1$, respectively. The sub-component density $u_m$ and distance $\zeta$ are also dimensionless, and we write the reduced GPEs (Eqs.\ref{eq6}-\ref{eq7}) in a dimensionless form for the stationary states. The total energy (scaled with $\hbar\omega_z$) is,
\begin{equation}\label{eq:energy}
    \begin{split}
        E=\int_0^\infty d\zeta \Bigg(&-\dfrac{1}{2}\sum_m \sqrt{u_m}\dfrac{d^2}{d\zeta^2}\sqrt{u_m} + \dfrac{1}{2}\zeta^2 u\\
        &- p'(u_1-u_{-1}) +q'(u_1+u_{-1})\\
        &+\dfrac{1}{2}\lambda_0 u^2+\dfrac{1}{2}\lambda_1(u_1-u_{-1})^2\\
        &+\lambda_1 u_0(u_1+u_{-1}+2\sqrt{u_1u_{-1}}\cos{\theta_r})\Bigg),
    \end{split}
\end{equation}
where $u$ is the total density and $p' = p/(\hbar\omega_z)$ and $q' = q/(\hbar\omega_z)$ are linear and quadratic Zeeman terms in dimensionless form.

Note that, in Eq.\ref{eq:energy}, the $E$ being integrated along the length of the confinement means it is the energy per unit cross-sectional area of the quasi-one-dimensional condensate. Now, since the lengths are scaled by the respective units $l_z$ and $l_{xy}$, the dimensionless lengths in all directions are equivalent. Taking into account the volume of the condensate being proportional to the total number of particles $N$, this makes the cross-sectional area scale as $N^{2/3}$. This scaling with particle number will be evident in the particle number dependence of the parameters $p^\prime$ and $q^\prime$ when plotting phase boundaries on the $p^\prime-q^\prime$ plane. This scaling can serve as a good cross-check for the correctness of the computations.

\par

The wavefunction for the different components of a stationary state is, in general, multi-modal in nature. We take an ansatz that the wavefunction (in dimensionless form) for the component with spin projection $m$ is described as, $\phi_m(\zeta)=(a_m - b_m\zeta^2)\exp\big(-\zeta^2/2d_m\big)$. The corresponding number density for $m^{th}$ sub-component is,
\begin{equation}\label{eq10}
u_m \equiv
|\phi_m|^2 = \Big[(a_m - b_m\zeta^2)\exp{\big(-\zeta^2/2d_m\big)}\Big]^2.
\end{equation}
Note that, in contrast to Ref.\cite{Kanjilal_PRA}, the wavefunction is taken as a single continuous function of $\zeta$, the distance from the centre of the trap. 
\par
Near the centre of the harmonic trap, where the potential energy resulting from trapping is the lowest, one can expect a higher density. Consequently, the combined effect of interaction energy ($\sim n^2(\zeta)$) and the potential energy ($\sim U(\zeta)n(\zeta)$) overshadows the effect of the kinetic energy. Thus, one can safely neglect the kinetic energy term to obtain the Thomas-Fermi approximated (TFA) profile of the condensate. 
\par
Naturally, we demand that the variational profile be as close as possible to the TFA profile near the centre of the trap for small values of $\zeta$. Away from the centre, the number density becomes lower, and so does the interaction. In this region, the Gaussian nature of the profile prevails because the linear terms in the GPEs dominate over the nonlinear ones, which is analogous to the harmonic oscillator wave function. Although the procedure is fairly general, for simplicity, let us take the AF state as an example to demonstrate the implementation of the proposed multi-modal variational method.  
\par
In the AF state, the sub-components corresponding to $m=0$ level is empty and the other two are populated. Following our ansatz in Eq.\ref{eq10}, we can write the sub-component densities as,
\begin{equation}\label{eq11}
        u_1(\zeta)=\left(a_1 - b_1\zeta^2\right)^2\exp{\left(-\frac{\zeta^2}{d}\right)},      
\end{equation}
\begin{equation}\label{eq12}
        u_{-1}(\zeta)=\left(a_{-1} - b_{-1}\zeta^2\right)^2\exp{\left(-\frac{\zeta^2}{d}\right)}.
\end{equation}

\par
We assume that the parameter $d_m$, which dictates the behaviour in the low-density region, for both sub-components to be equal, i.e., $d_1=d_{-1}=d$, by symmetry (although, in reality, that remains weakly broken for such systems).
Under this assumption, one gets to the total number density of the AF state near the centre of the condensate as,
\begin{equation}\label{eq13}
    \begin{split}
        u(\zeta)=\Big[(a_1^2&+a_{-1}^2)+(b_1^2+b_{-1}^2)\zeta^4\\
        &-(2a_1b_1+2a_{-1}b_{-1})\zeta^2\Big]\exp\bigg(-\frac{\zeta^2}{d}\bigg),
    \end{split}
\end{equation}
which, in total, has five unknown parameters: $a_1,a_{-1},b_1,b_{-1},d$, and the chemical potential $\mu'$, which is to be estimated by these parameters and the number of particles in the system.

In TFA, one simplifies the GPEs by neglecting the kinetic energy, and the corresponding solution (of the wavefunction or the number density) is referred to as the TFA profile. As noted earlier, close to the trap centre, the interaction energy contribution is much more significant than the kinetic energy. Therefore, in this regime, one can expect the TFA profile to remain relatively accurate. Hence, we equate the number density (Eq.\ref{eq13}) with the TFA profile. For small enough $\zeta$, expanding the exponential term and neglecting the higher-order terms $\mathcal{O}(\zeta^4)$, one can estimate the total density,
\begin{equation}\label{eq14}
    u(\zeta)=(a_1^2+a_{-1}^2)-\zeta^2\left(\frac{a_1^2}{d}+\frac{a_{-1}^2}{d}+2a_1b_1+2a_{-1}b_{-1}\right),
\end{equation}
which should match the corresponding TFA profile for this state \cite*{Kanjilal_2020,Kanjilal_corri},
\begin{equation}\label{eq15}
    u^{TFA}_{tot}=\dfrac{\mu'-q'-\zeta^2/2}{\lambda_0}.    
\end{equation}
\par
Equating the expression corresponding to the total number densities (Eqs.\ref{eq14}-\ref{eq15}), leads to the relations between the six unknown quantities, where
\begin{subequations}\label{eq16}
    \begin{equation}\label{eq16a}
    \mu'=\lambda_0(a_1^2+a_{-1}^2)+q'
    \end{equation}
    \begin{equation}\label{eq16b}
    b_1=\dfrac{\Big(\dfrac{1}{2\lambda_0}-\dfrac{a_1^2}{d}-\dfrac{a_{-1}^2}{d}-2a_{-1}b_{-1}\Big)}{2a_1}.
    \end{equation}
\end{subequations}
These two relations leave us, at this stage, with four unknown quantities, namely $a_1,\ a_{-1},\ b_{-1}$ and $d$. 

\par
Our next step is to minimise the free energy of our system with respect to these parameters, and to use the number conservation equation as a constraint to determine them. We employ the Lagrange multiplier method in the presence of a constraint, i.e,
\begin{subequations}\label{eq17}
    \begin{equation}
        N-\left(\int u_{-1}(\zeta) \, d\zeta + \int u_{1}(\zeta) \, d\zeta\right) =0,
    \end{equation}
    \begin{equation}
        \nabla_{a_1}\bigg[E-\mu' \left(\int u_{-1}(\zeta) \, d\zeta + \int u_{1}(\zeta) \, d\zeta\right)\bigg]=0,
    \end{equation}
    \begin{equation}
        \nabla_{a_{-1}}\bigg[E-\mu' \left(\int u_{-1}(\zeta) \, d\zeta + \int u_{1}(\zeta) \, d\zeta\right)\bigg]=0,
    \end{equation}
    \begin{equation}
        \nabla_{b_{-1}}\bigg[E-\mu' \left(\int u_{-1}(\zeta) \, d\zeta + \int u_{1}(\zeta) \, d\zeta\right)\bigg]=0,
    \end{equation}
\end{subequations}    
which fixes the remaining unknown parameters. Here, $E$ represents the energy (Eq.\ref{eq:energy}), and $N$ is the total number of particles present in the condensate. The final step is to solve these four equations, which simultaneously fix the remaining four unknown parameters. The variational parameters determine the condensate profile, which in turn gives the energy of the system for a particular value of $q',\ p'$, and $N$.
\par
Note that this method is much simpler than the variation procedure introduced earlier \cite{kanjilal_variational, Kanjilal_PRA} by some of us. The procedure used here, along with the choice of a single continuous function as the condensate profile, greatly simplifies the estimation of variational parameters. Together with Eq.\ref{eq16}, after integrating over the spatial parts, Eq.\ref{eq17} reduces to a set of nonlinear algebraic equations that can be solved (using the Newton-Raphson method) to determine the parameters. 
\par
Note that the variational method presented here is discussed in the context of the AF state. The same method for the other multicomponent state, namely the PM state, is discussed in Appendix A. For the single-component states, such as ferromagnetic and polar states, the method can be applied straightforwardly.

\subsection{The comparison of the variational and numerical results of the antiferromagnetic state}
We consider a spin-1 $^{23}Na$ condensate for which the spin interaction is antiferromagnetic in nature. For comparison of the condensate profile obtained from the variational method with the numerical simulation, we consider a quasi-one-dimensional harmonic trapping geometry with oscillator lengths along the transverse direction, $l_{yz}=0.59\mu m$, and that along the longitudinal direction, $l_x=2.97\mu m$. This corresponds to $\lambda_1=7.43 \times 10^{-4}$ and $\lambda_0=46.16 \times 10^{-3}$, which are the spin-spin and spin-independent interaction, respectively. We consider a condensate of 5000 particles with fixed Zeeman terms, $q'=-0.5$ and $p'=0.2$. In this parameter regime, the AF state is favoured energetically.

The variational parameters determined from the solution of the Eq.\ref{eq16}-\ref{eq17} are $a_1\simeq24.95$, $a_{-1}\simeq16.52$, $b_1=-0.57$, $b_{-1}\simeq-0.37$ , $d\simeq 17.38$ and $\mu'\simeq 40.83$, leading to the analytical form of the number densities,
\begin{equation}
       u_1(\zeta)=\left(24.95 +0.57\zeta^2\right)^2\exp{\left(-\frac{\zeta^2}{17.38}\right)} ,
\end{equation}
\begin{equation}
       u_{-1}(\zeta)=\left(16.52 +0.37\zeta^2\right)^2\exp{\left(-\frac{\zeta^2}{17.38}\right)} .
\end{equation}

\begin{figure}[h]
    \centering
    \subfloat[]{
    \includegraphics[width=0.8\linewidth]{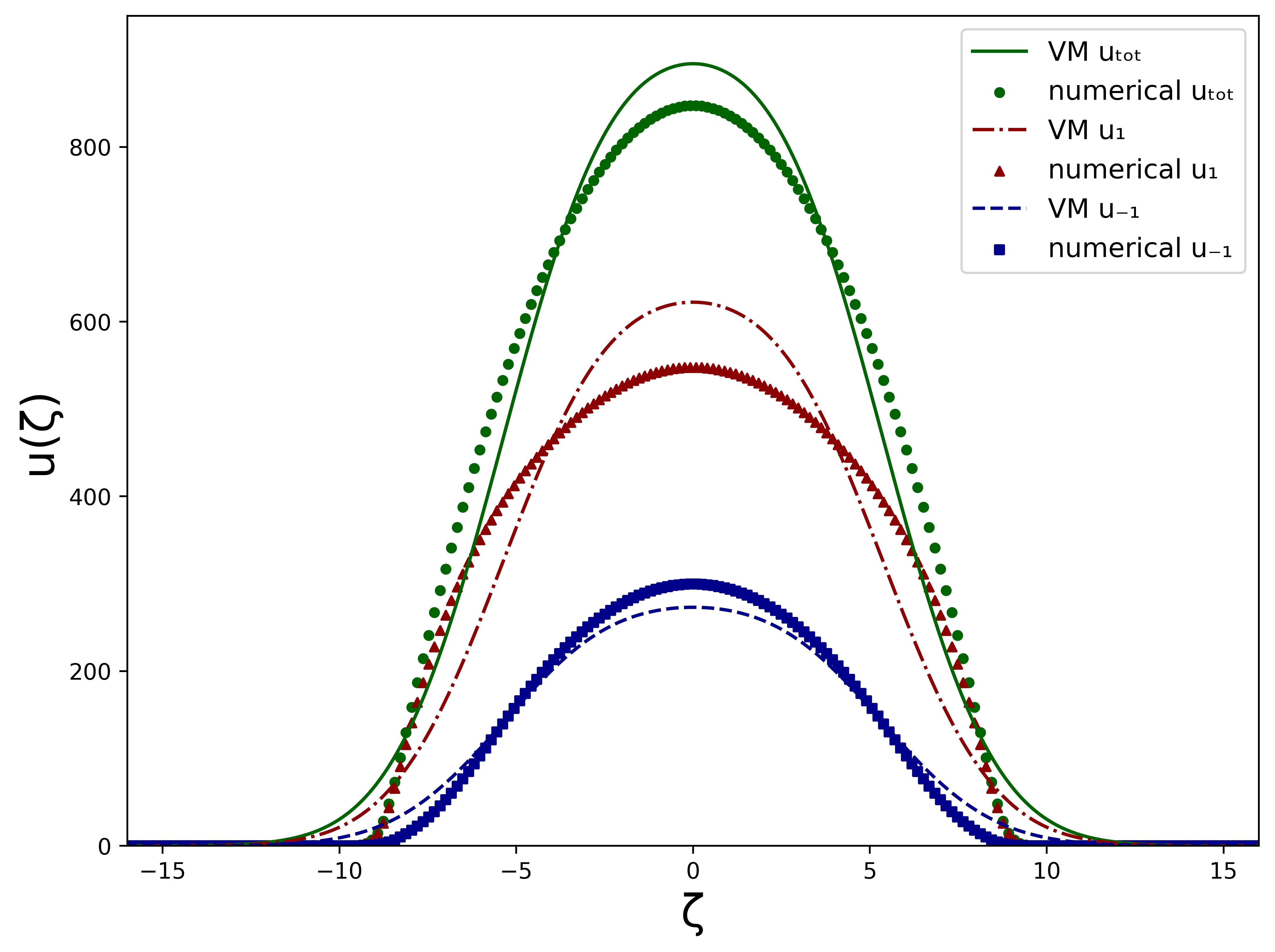}\label{subfig:2a}
    }
    \\
    \subfloat[]{
    \includegraphics[width=0.8\linewidth]{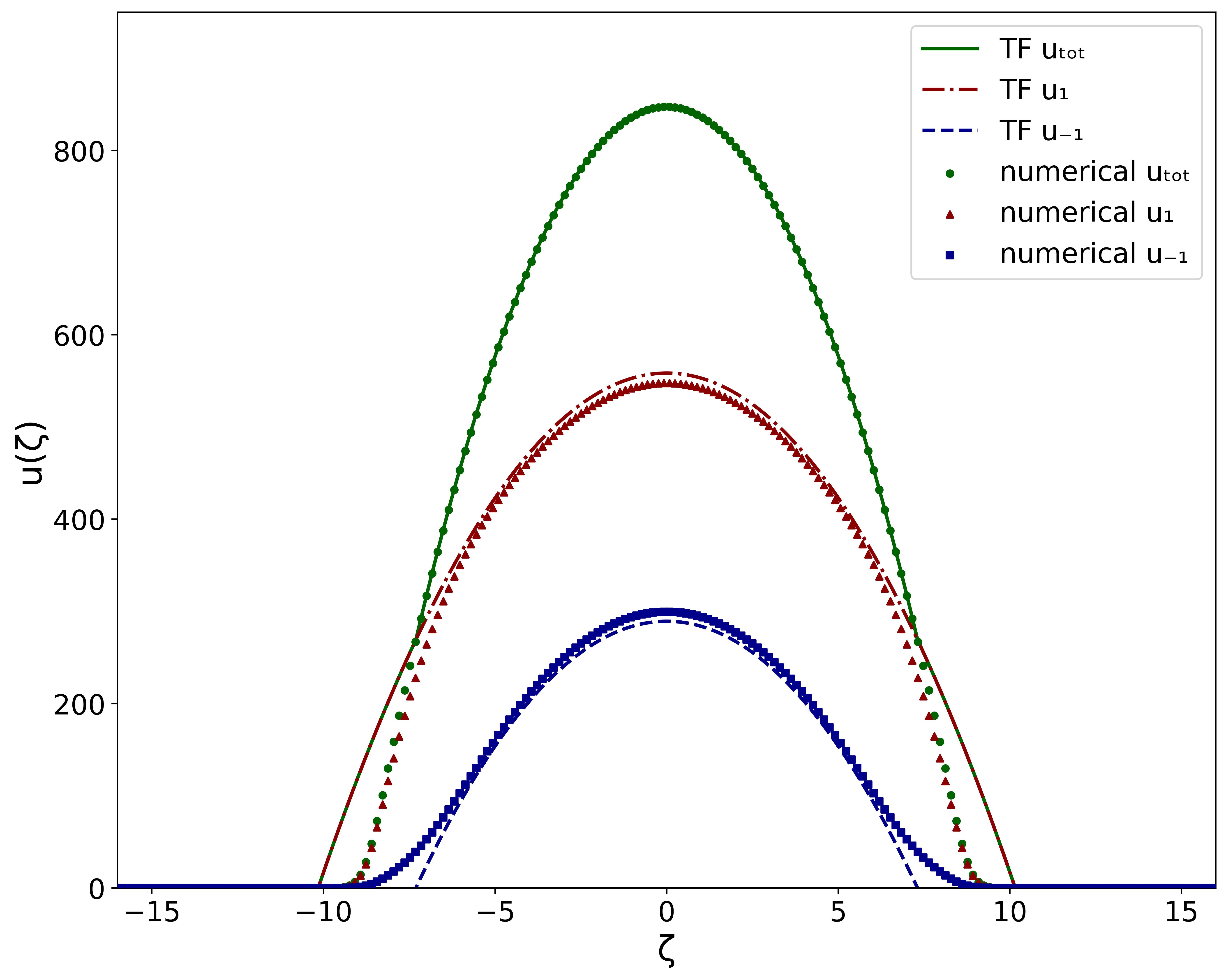}\label{subfig:2b}
    }
    \caption{The total density $u_{tot}$ and sub-component densities $u_1$ and $u_{-1}$ as a function of the distance $\zeta$ from the centre of the quasi-1D harmonic trapping, obtained from the (a) variational method, (b) Thomas-Fermi approximation (in green-solid, red-dash-dot and blue-dashed line respectively), are compared against the same obtained from numerical simulation (in green-circle, red-triangle and blue-square respectively) for the antiferromagnetic state at $p'=0.2$ and $q'=-0.5$.}
    \label{fig:2}
\end{figure}

In Fig.\ref{fig:2}\subref{subfig:2a}, we compare the variationally obtained number density with the numerically obtained profiles. We employ the imaginary-time split-step Fourier method \cite{gautam_GPE_solver} to solve the GPE for the parameters specified earlier numerically.

In Fig.\ref{fig:2}\subref{subfig:2b}, the TFA condensate profile is compared with the numerical profile. The total density from TFA (in green line) roughly matches that obtained from the numerical method (green circle) near the centre of the trap, and this gives a false impression that it may be better to use the TFA, in general, for further energy estimation of the state. The main issue with TFA first appears in the low-density region. The sub-component density from TFA (in this case $u_{-1}$ shown in blue-dash line) sharply goes to zero at the Thomas-Fermi radius, in contrast to the expected smoothly varying profile as obtained in both variational and numerical methods. Secondly, beyond the Thomas-Fermi radius of one of the components, in this case, $u_{-1}$ (shown in a blue dashed line), the other component(s), in this case, $u_1$ (shown in a red dash-dot line), overshoot the numerical estimate. Since $u_{-1}$ is zero in this region, the total density in TFA (shown in the green line) also overshoots the numerical estimate. Another argument is that when one of the components goes to zero, the state disappears, and a domain forms with another state. In this case, when the $u_1$ is only present beyond the Thomas-Fermi radius of the $u_{-1}$, one can argue according to TFA that there is a domain of AF state near the centre of the trap and a ferromagnetic state F1 beyond the Thomas-Fermi radius of $u_{-1}$. 
\par
Note that the numerical results do not suggest that there are any such domain-forming states, and that is the precise reason the TFA fails for multi-component states, which is discussed in more detail in \cite{Kanjilal_PRA}. The shortcomings of TFA necessitate an analytical method, which is where the proposed variational method comes in.

\section{Estimating Phase Boundaries from a Variational Method: A Route to the Complete Phase Diagram}

Estimating the variational parameters for a particular state also determines the system's energy. The energies of different states can be compared to get the phase boundary. In the following, we demonstrate the importance of the variational method in obtaining the complete phase diagram of the system. We first focus on the case of condensates with antiferromagnetic interactions, and then we move on to the case of ferromagnetic interactions.

\subsection{Phase boundary for Condensates having $\lambda_1>0$ under harmonic confinement}\label{pbd_c1>0}

For a homogeneous system with antiferromagnetic type interaction, the ferromagnetic states, polar, and AF states appear as the ground state in different parameter regimes, as shown in Fig.\ref{fig:1}\subref{fig:1a}. The phase boundary between the AF and polar phase is parabolic in $p$, $q$ parameter space, while the ferromagnetic-polar phase boundary is linear. Note that the AF-ferromagnetic phase boundary does not depend on $q$ for the homogeneous case, and is determined by the vanishing of one of the sub-component densities (either $n_1$ or $n_{-1}$) of the AF state, which happens at $|p|=c_1 n$. 
\par
In the presence of trapping, there is no guarantee that the nature of these phase boundaries will remain entirely unaltered. This is because of the fact that the number density, which is uniform in the case of a homogeneous condensate, does not remain so under trapping. On top of that, the condensate profiles are multi-modal in nature, so the effects of trapping and the ensuing kinetic energy contribution can play a crucial role, as captured in a simplified manner by the variational method introduced here.

\subsubsection{Antiferromagnetic(AF)-Polar phase boundary}
Comparing the variationally estimated total energy of the AF state and the polar state would yield the phase boundary for different values of $q^\prime$ and $p^\prime$. This comparison is done in quasi-1D harmonic confinement with spin-independent interaction fixed at $\lambda_0= 46.16\times10^{-3}$. 

\begin{figure*}[htb]
  \centering
  \subfloat[]{
    \includegraphics[width=0.32\linewidth]{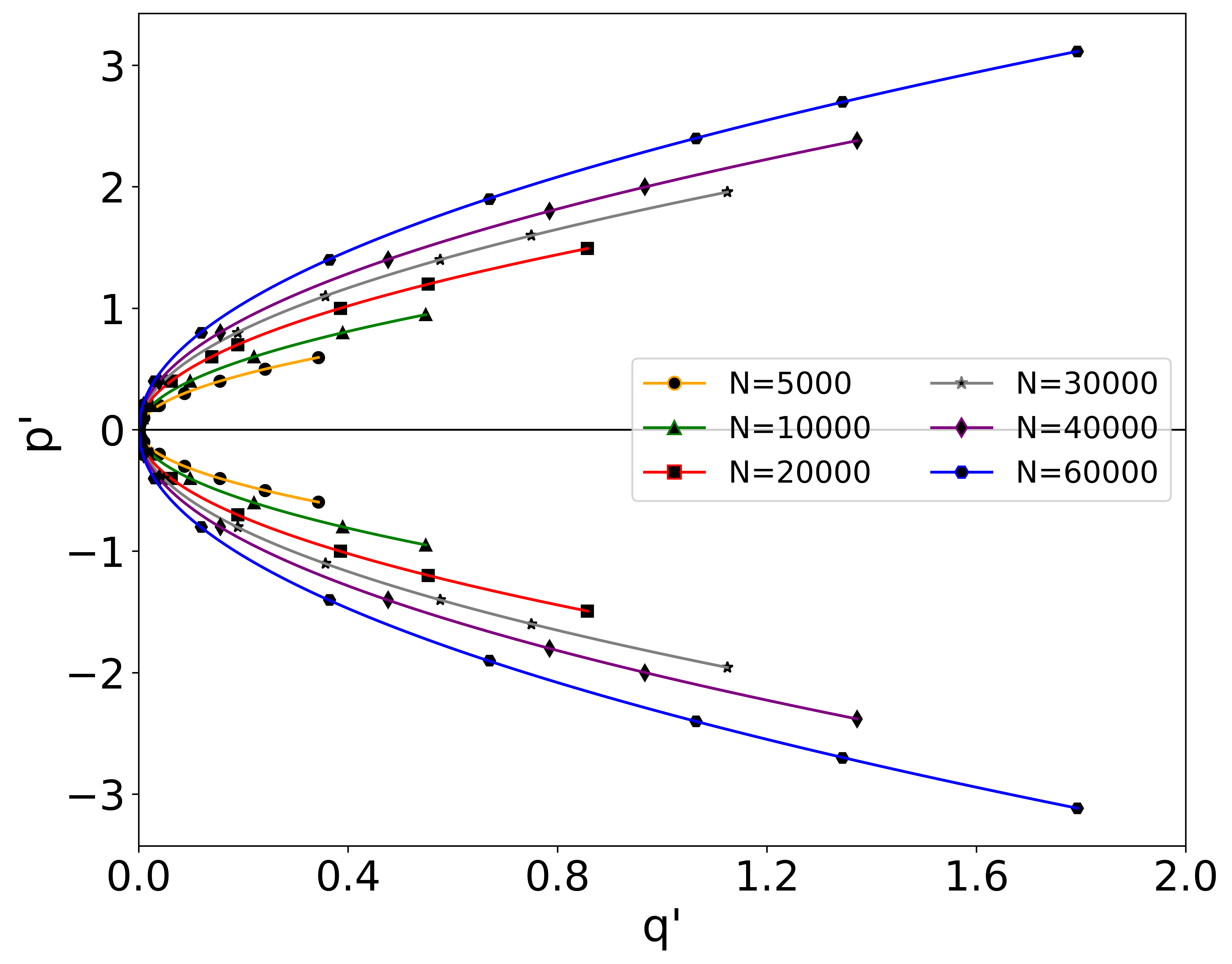}
    \label{fig:3a}
  }
  \subfloat[]{
    \includegraphics[width=0.32\linewidth]{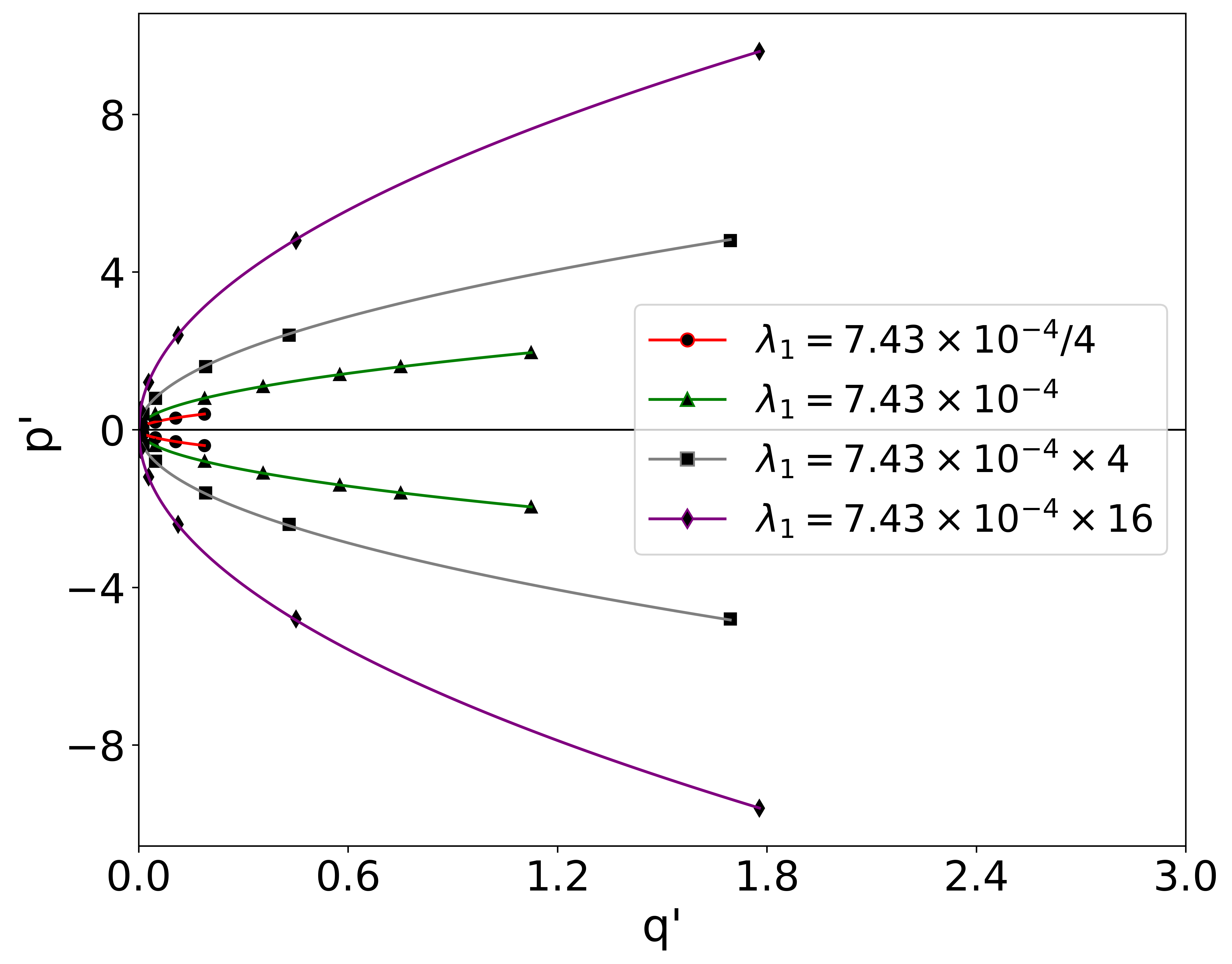}
    \label{fig:3b}
  }
  \subfloat[]{
    \includegraphics[width=0.32\linewidth]{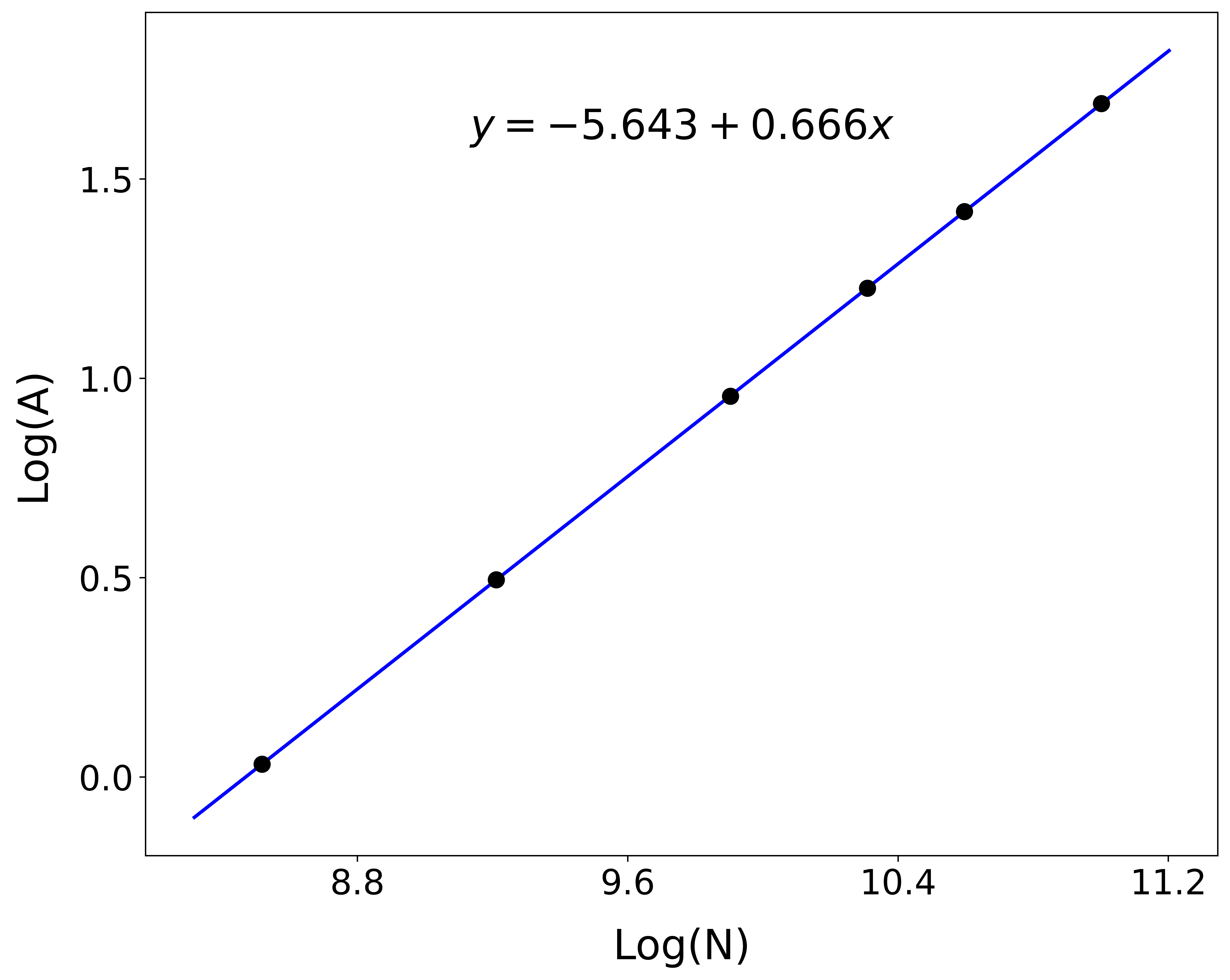}
    \label{fig:3c}
  }
  \\
  \subfloat[]{
    \includegraphics[width=0.32\linewidth]{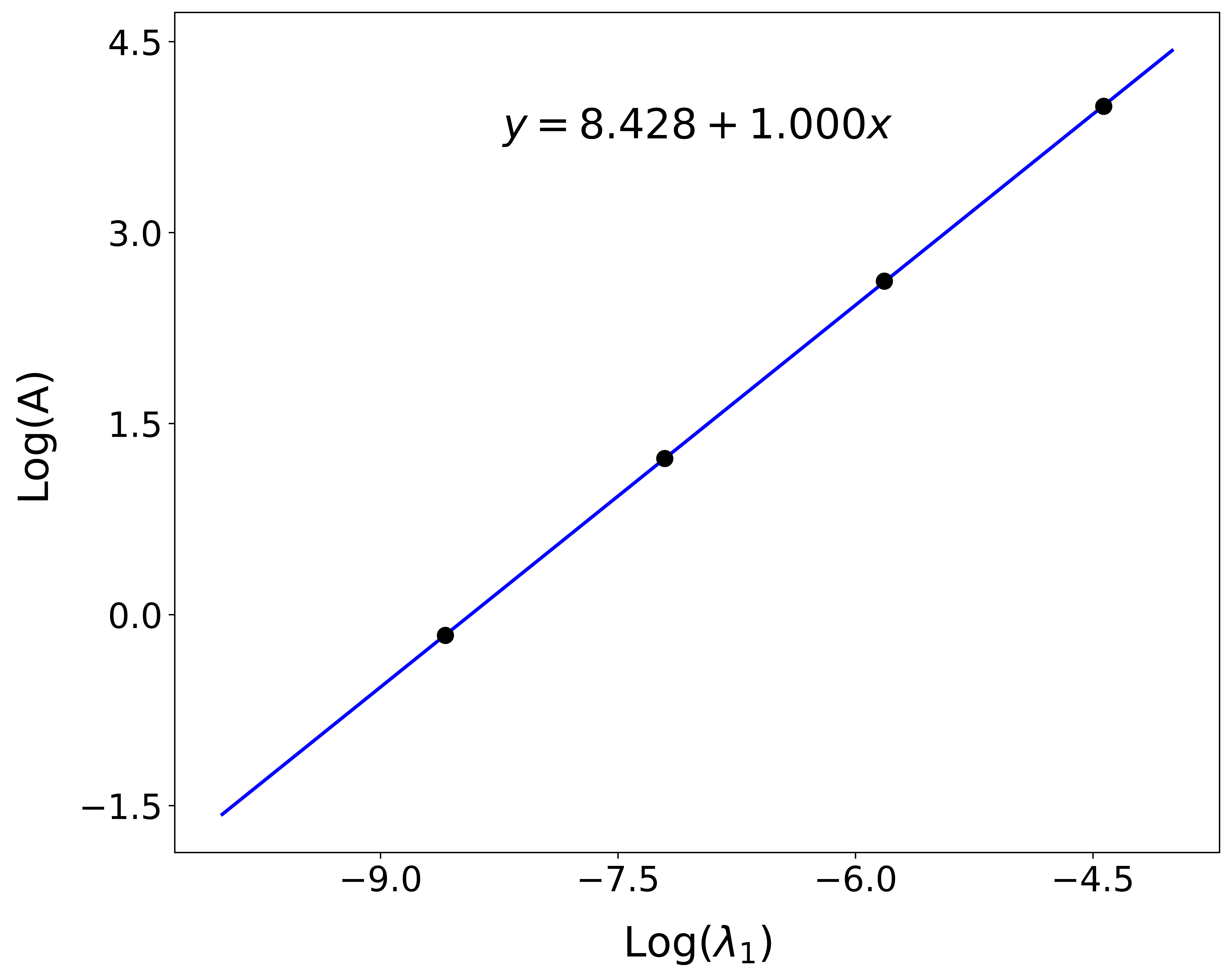}
    \label{fig:3d}
  }
  \subfloat[]{
    \includegraphics[width=0.32\linewidth]{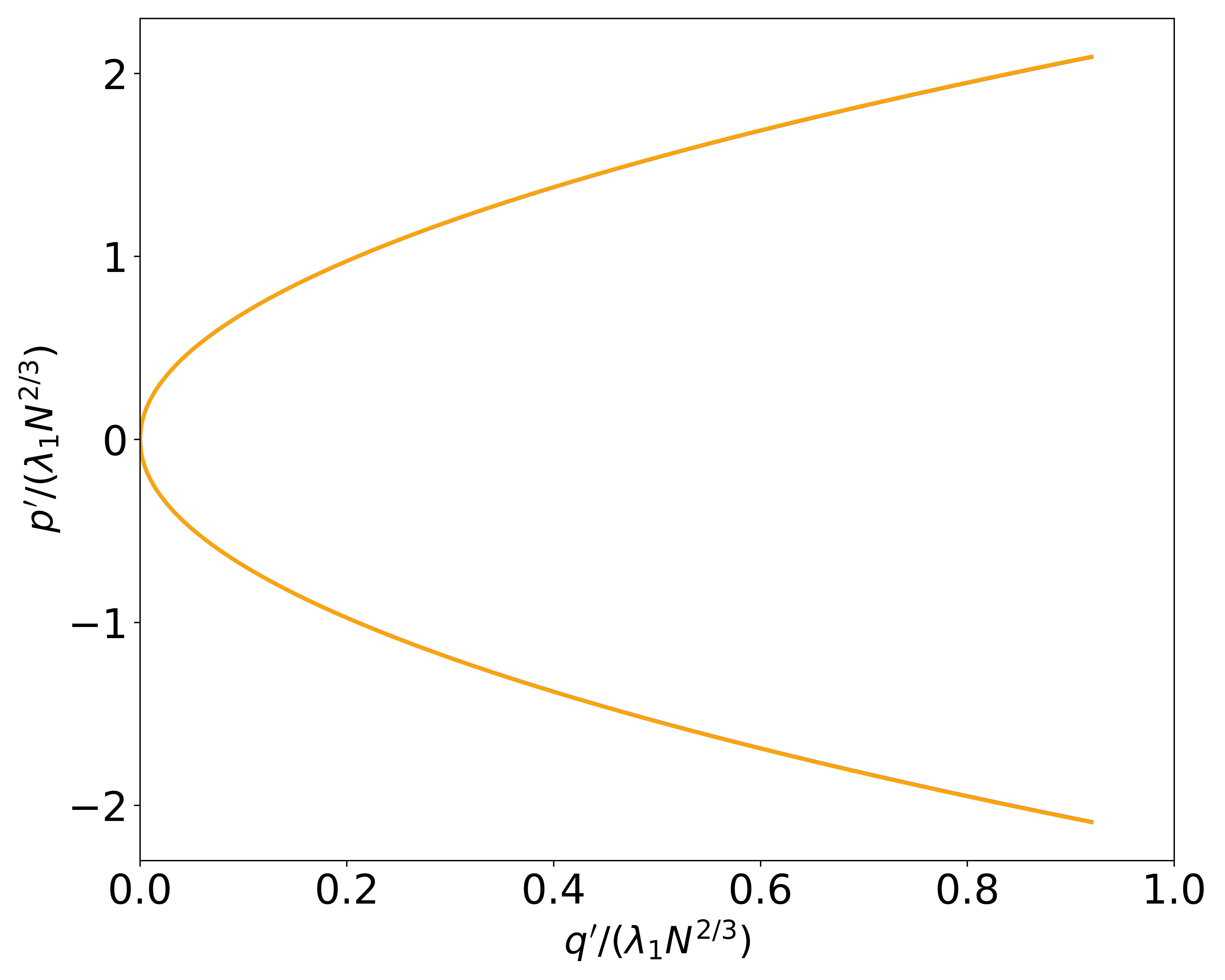}
    \label{fig:3e}
  }
  \caption{(a) Phase boundary between anti-ferromagnetic and polar state with varying condensate particles of a spin-1 condensate with anti-ferromagnetic type of spin interaction. (b) The same type of phase boundary with varying spin-interaction strength for a fixed condensate particle of $N=30000$. (c) The scaling factor $A(N,\lambda_1)$ scales the phase boundary of the anti-ferromagnetic state and the polar state as $p'^2 = A(N,\lambda_1) q'$. The variation of scaling factor $A(N,\lambda_1)$ with the number of condensate particles $N$ and (d) with the spin-dependent interaction $\lambda_1$. (e) The universal phase boundary obtained after scaling the linear and quadratic Zeeman terms with the scaling factor $\lambda_1 N^{2/3}$.}
  \label{fig:3}
\end{figure*}

In Fig.\ref{fig:3}\subref{fig:3a}, we plot the phase boundaries between the AF and the polar state for varying condensate size, i.e., different total number of particles $N$ and for a fixed $\lambda_1= 7.43 \times 10^{-4}$. Even in the presence of trapping, we find that the boundaries behave like a system of parabolas, $p'^2=A(N,\lambda_1) q'$, with $A$ being a function of $N$ ($\lambda_1$ is fixed in this case). 
\par
The factor $A(N,\lambda_1)$ also depends on the spin-spin interaction $\lambda_1$. To capture that, we calculate the phase boundary by fixing the particle number $N = 30000$ and varying $\lambda_1$, to $1/4$, $4$ and $16$ times of the previously fixed value (Fig.\ref{fig:3}\subref{fig:3b}). Note that the spin-spin interaction can be varied experimentally as well, using the Feshbach resonance \cite{Feshbach}. 
\par
From the collapse of the phase boundaries in Fig.\ref{fig:3}\subref{fig:3a}, we obtain the fitting function $A(N,\lambda_1)$ for different plots corresponding to different $N$. We assume a power-law dependence of the fitting parameter $A(N,\lambda_1)$ on $N$. We find $A(N,\lambda_1)\propto N^{2/3}$, for a fixed value of $\lambda_1$ (see Fig.\ref{fig:3}\subref{fig:3c}). Following a similar approach of collapsing the boundaries in Fig.\ref{fig:3}\subref{fig:3b}, we find the fitting parameter $A(N,\lambda_1)$ for different $\lambda_1$ for a fixed $N$. In Fig.\ref{fig:3}\subref{fig:3d}, we find $A(N,\lambda_1)\propto \lambda_1$.

Now, we can scale the $p'$ and $q'$ axes with $N^{2/3}\lambda_1$, which are the controllable parameters. This produces a universal feature of the phase boundary across the wide range of parameters considered (shown in Fig. \ref{fig:3}\subref{fig:3e}). Interestingly, the qualitative prediction of the homogeneous case, i.e., the parabolic nature of the AF-Polar phase boundary, persists even in the case of a harmonically trapped quasi-1D spin-1 condensate.

\subsubsection{Ferromagnetic-Polar phase boundary}
A similar comparison can be made to determine the phase boundary between the polar and ferromagnetic states. As shown in the homogeneous phase diagram (Fig.\ref{fig:1}\subref{fig:1a}), there are two possible ferromagnetic states, F1 and F2, both of which have phase boundaries with polar states for $p>0$ and $p<0$, respectively. As both cases are symmetric, we focus only on obtaining the phase boundary between the F1 state and the polar state at $p>0$. 

In Fig.\ref{fig:4}\subref{fig:4a}, the phase boundaries are calculated for different numbers of particles and fitting suggests that they follow a linear dependence $p'=q'+B$, where $B$ is the intercept. We find that $B$ follows the same scaling with respect to $N$ and $\lambda_1$, i.e., $B\propto  N^{2/3}\lambda_1$.

\subsubsection{Combining the phase boundaries: Phase diagram for $\lambda_1>0$}
\begin{figure*}[htb]
    \centering
    \subfloat[]{\includegraphics[width=0.32\linewidth]{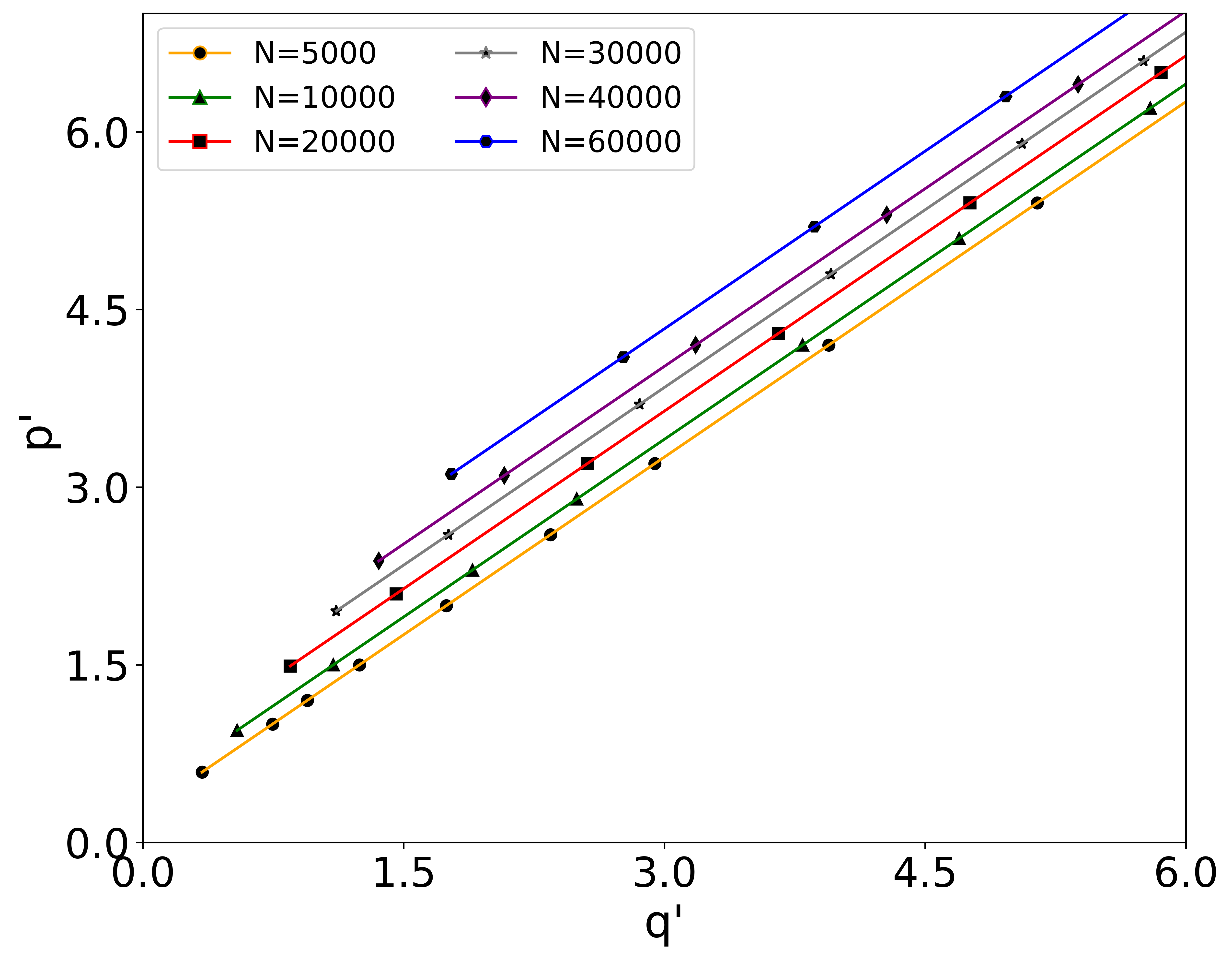}
    \label{fig:4a}
    }
    \subfloat[]{\includegraphics[width=0.32\linewidth]{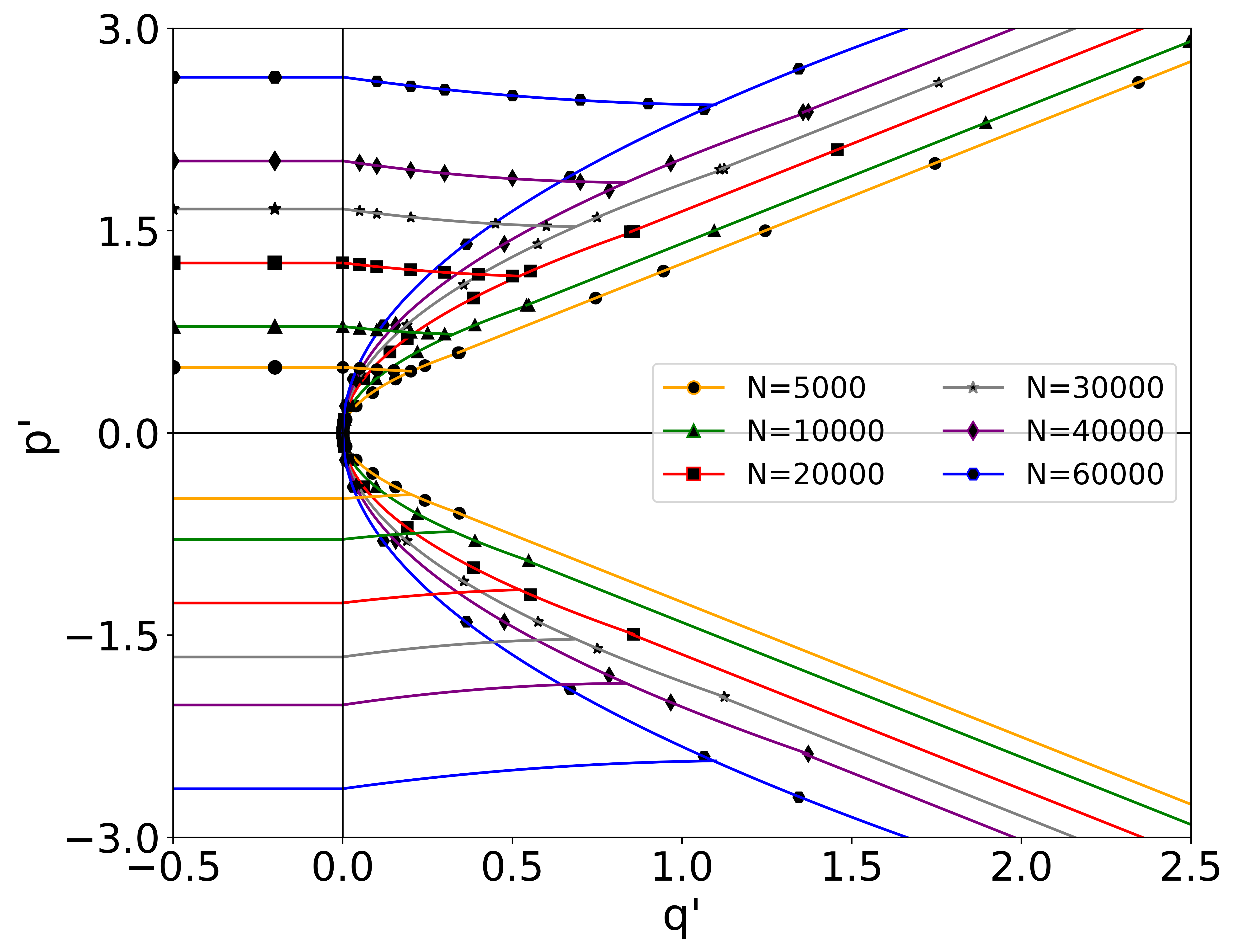}
    \label{fig:4b}
    }
    \subfloat[]{\includegraphics[width=0.32\linewidth]{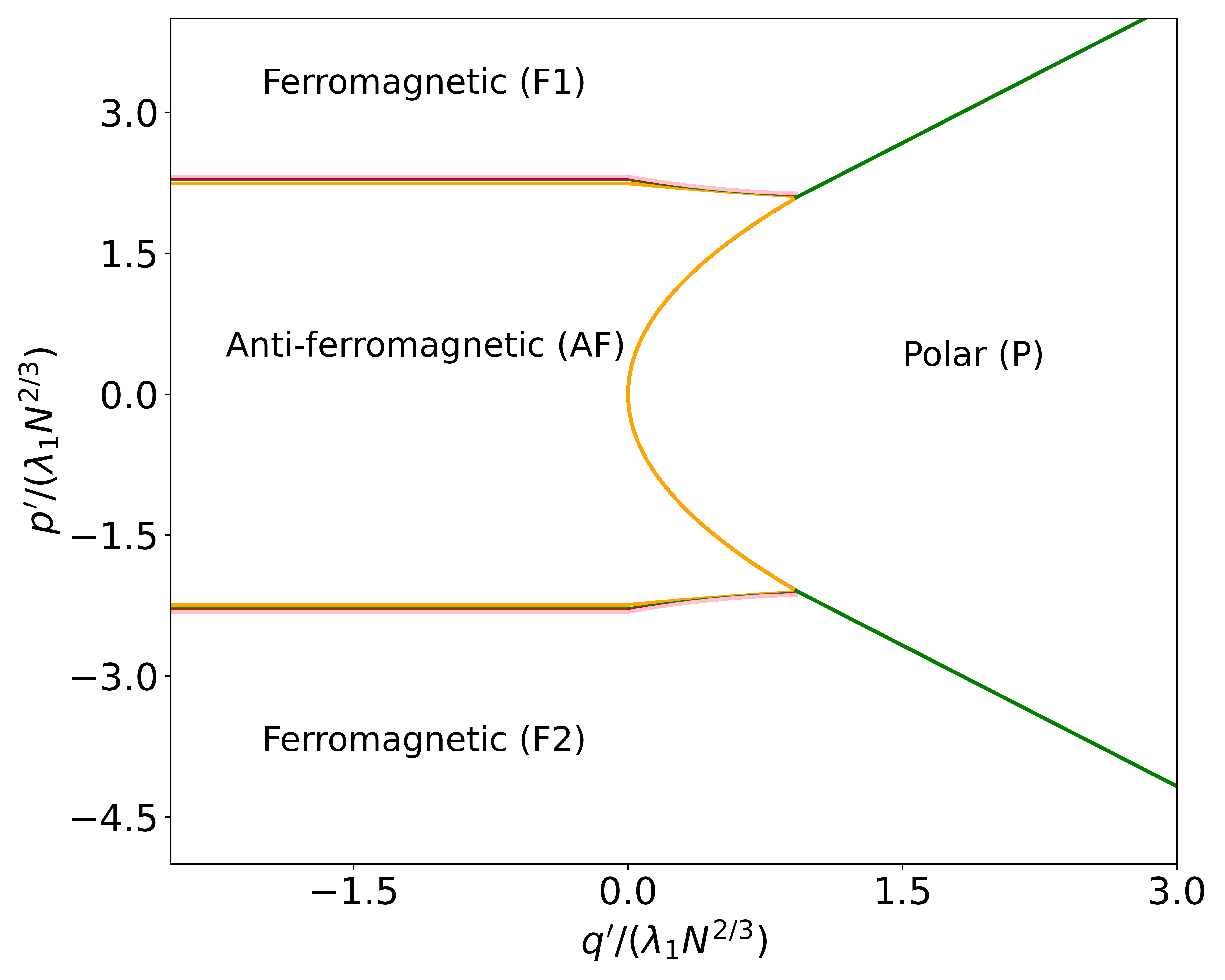}
    \label{fig:4c}
    }
    \caption{(a) The phase boundary between Ferromagnetic (F1) and polar states for different $N$ values. (b) Complete phase diagram with $\lambda_1>0$ for trapped condensates with varying values of $N$ in $(q',\:p')$ parameter space. (c) Universal phase diagram for all $N$ and $\lambda_1$ after scaling the linear and quadratic Zeeman strengths.}
    \label{fig:4}
\end{figure*}

We have already determined the phase boundaries and associated scaling behaviour of the AF-polar and ferromagnetic-polar phase boundaries. We include the phase boundaries between the ferromagnetic and anti-ferromagnetic states to get the complete phase diagram (see Fig.\ref{fig:4}\subref{fig:4b}). Upon the scaling of $p'$ and $q'$ with $\lambda_1 N^{2/3}$, we get an universal behavior as depicted in Fig.\ref{fig:4}\subref{fig:4c}.

\par
Note that for negative values of $q'$, the phase boundary between ferromagnetic and antiferromagnetic states is independent of $q'$, which happens as one of the sub-component densities of the antiferromagnetic state becomes zero at the phase boundary, effectively becoming the ferromagnetic state. An interesting feature we found is that for $q'>0$, there is a slight slope in the same phase boundary, which is due to the anti-ferromagnetic state becoming a higher energy state in comparison to the ferromagnetic states, while both of the sub-components of the AF state are still non-zero. This particular feature, predicted by the variational method, is absent in the homogeneous case. 

\subsection{Phase boundary for Condensates having $\lambda_1<0$ under harmonic confinement}\label{pbd_c1<0}

Another interesting situation arises when the spin-spin interaction is ferromagnetic, allowing other stationary states to become the ground state in certain parameter regimes. This results in a different phase diagram, as shown for the homogeneous case in Fig.\ref{fig:1}\subref{fig:1c}. For positive $q$, with $q\gtrsim |p|$, the phase-matched state is energetically favored, followed by the polar state at even higher quadratic Zeeman strength. In the negative $q$, and positive $p$, the ferromagnetic F1 state dominates. The phase diagram is overall symmetric if we change the sign of $p$, the only difference is in place of F1, the other spin-reversed ferromagnetic state F2 appears, which is natural.

For the trapped condensate with ferromagnetic type of spin-interaction, we take the example of spin-1 ${}^{87}\text{Rb}$ condensate (which has $\lambda_1<0$) in a quasi 1-D harmonic confinement with oscillator lengths, $l_{zy}=0.30\mu m$, and $l_z=1.53\mu m$. The interaction parameters are $\lambda_0=17.66 \times 10^{-2}$, and $\lambda_1=-6.22 \times 10^{-4}$. Following a similar approach, we employ the variational method to investigate the various possible phase boundaries, thereby obtaining the complete diagram.

\begin{figure*}[htb]
    \centering
    \subfloat[]{
    \includegraphics[width=0.32\linewidth]{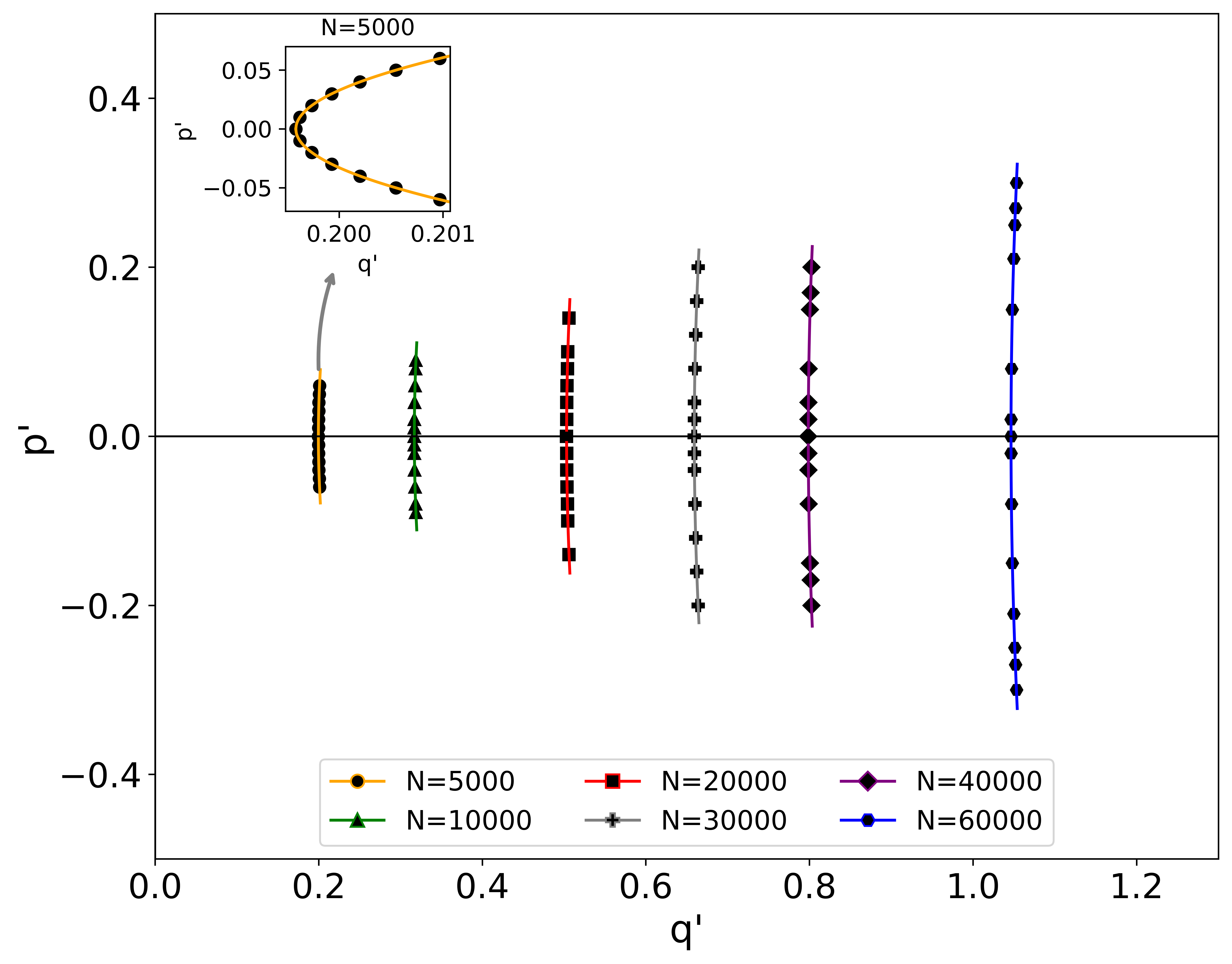}
    \label{fig:5a}
    }
    \subfloat[]{
    \includegraphics[width=0.32\linewidth]{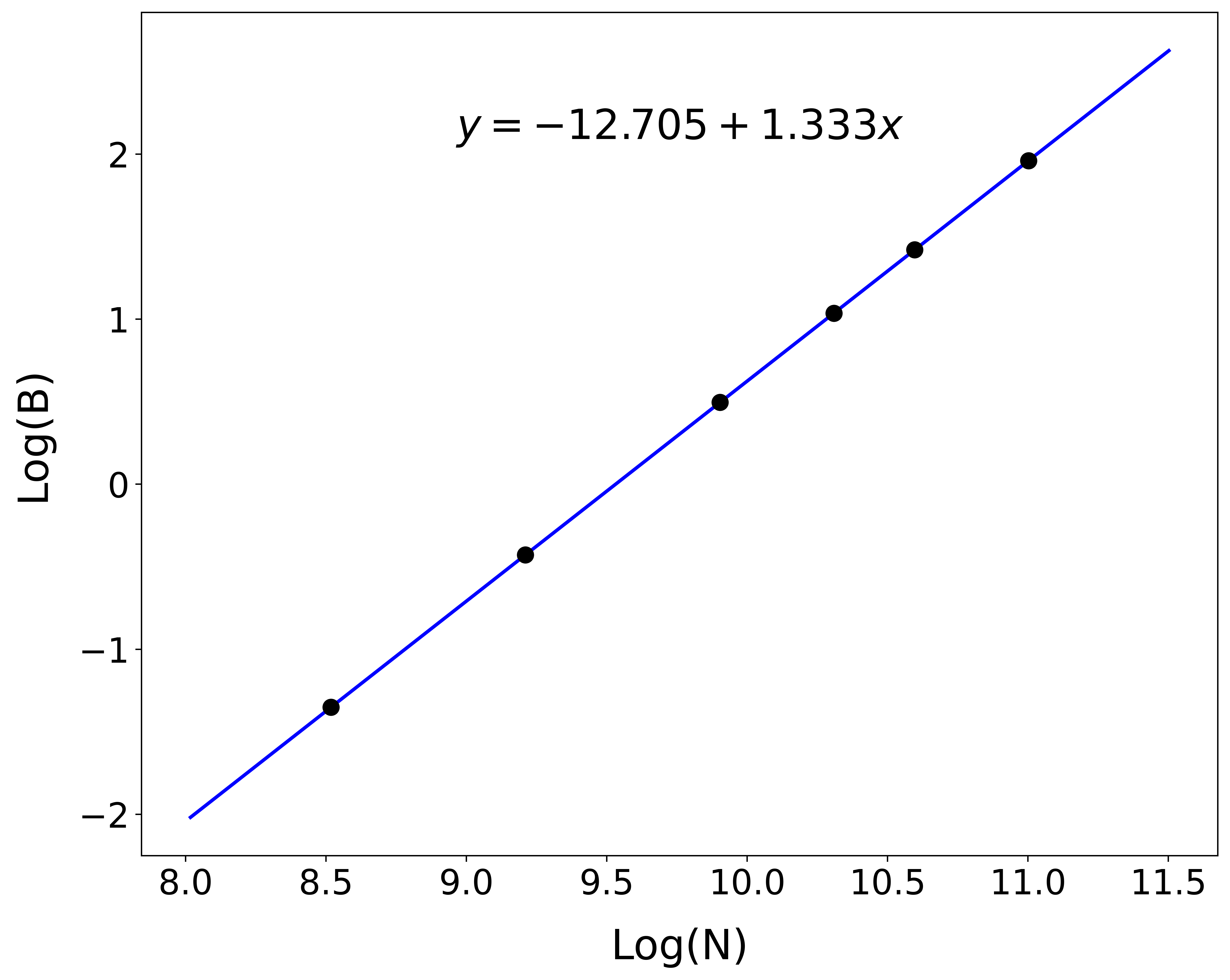}
    \label{fig:5b}
    }
    \subfloat[]{
    \includegraphics[width=0.32\linewidth]{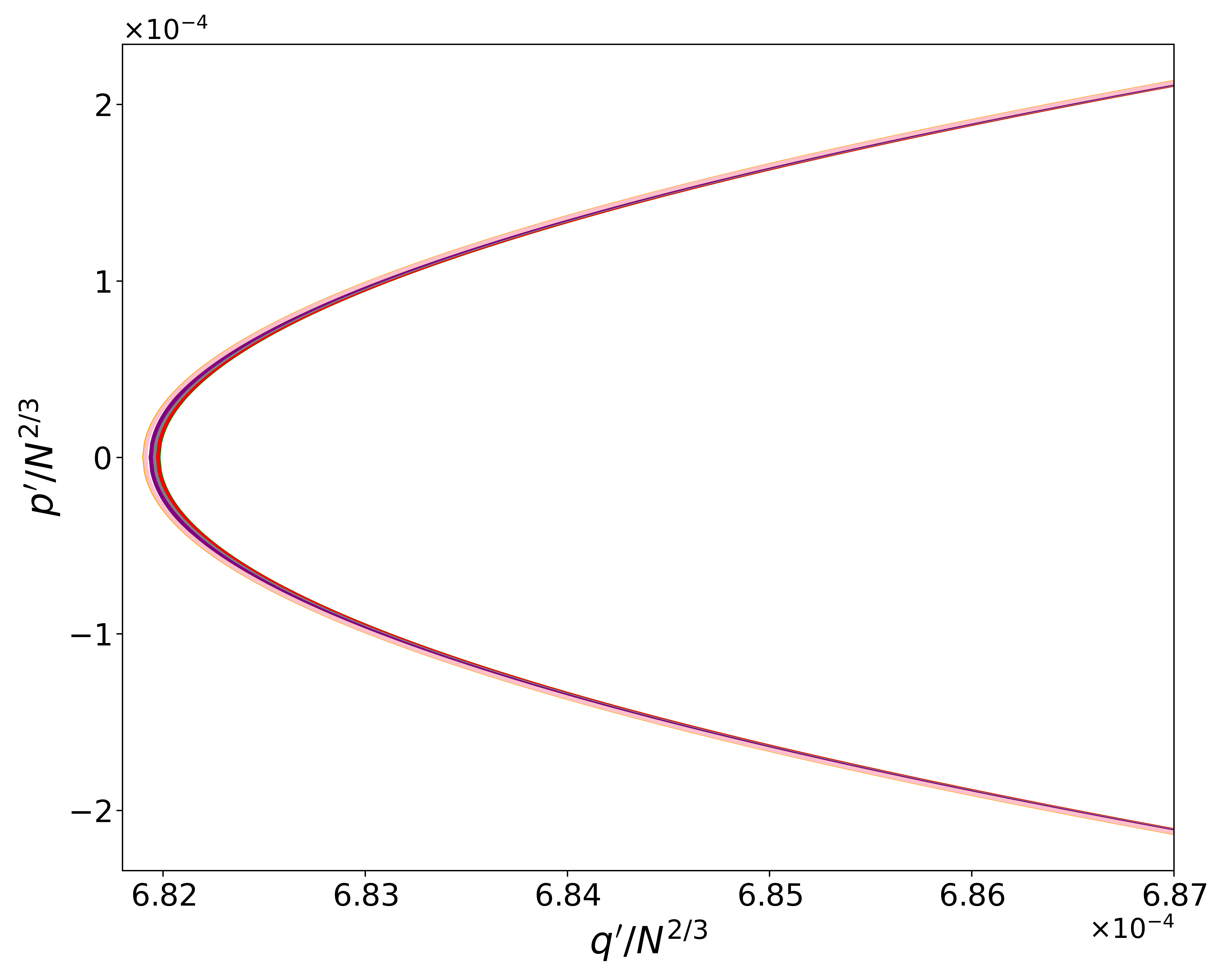}
    \label{fig:5c}
    }
    \\
    \subfloat[]{
    \includegraphics[width=0.32\linewidth]{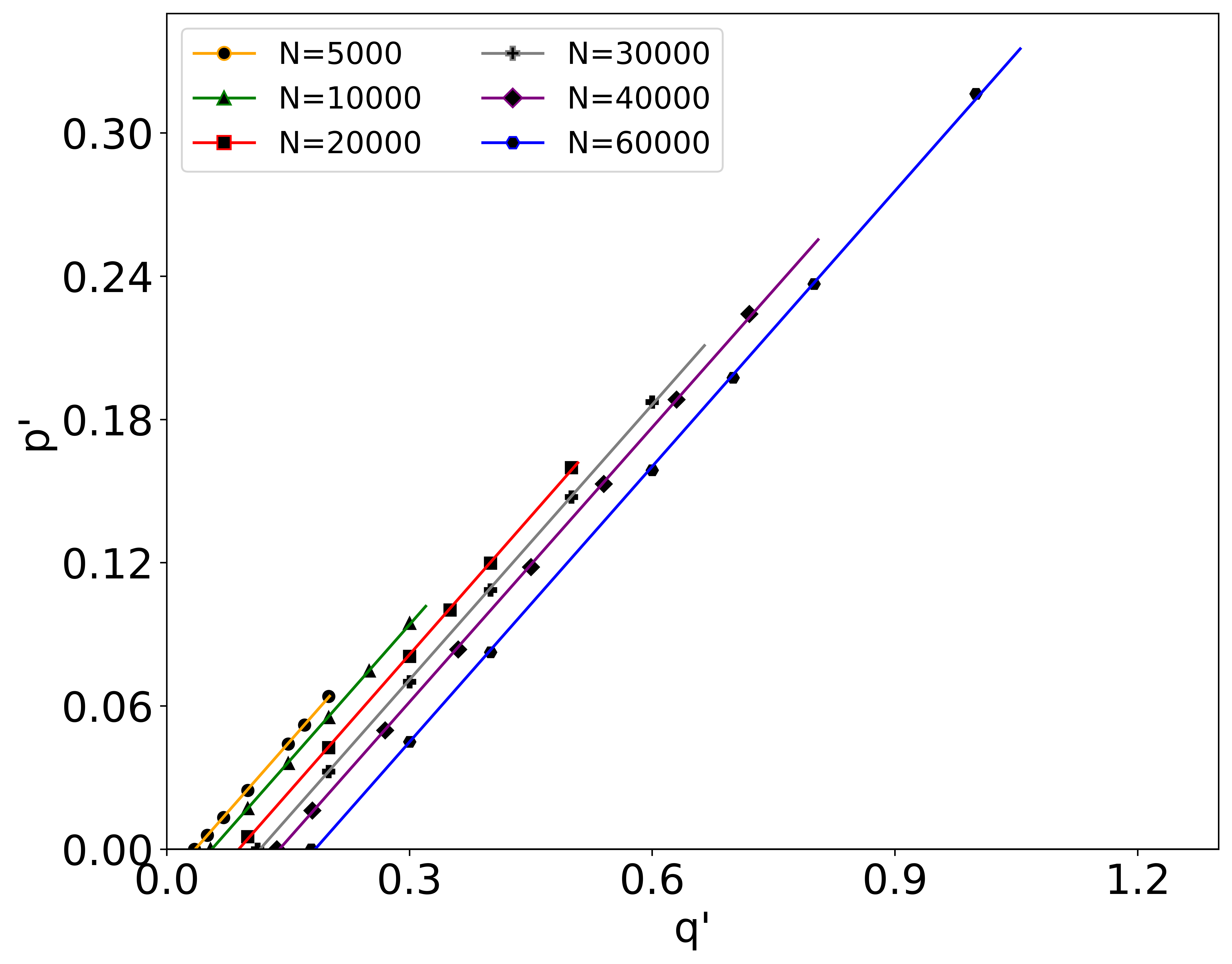}
    \label{fig:5d}
    }
    \subfloat[]{
    \includegraphics[width=0.32\linewidth]{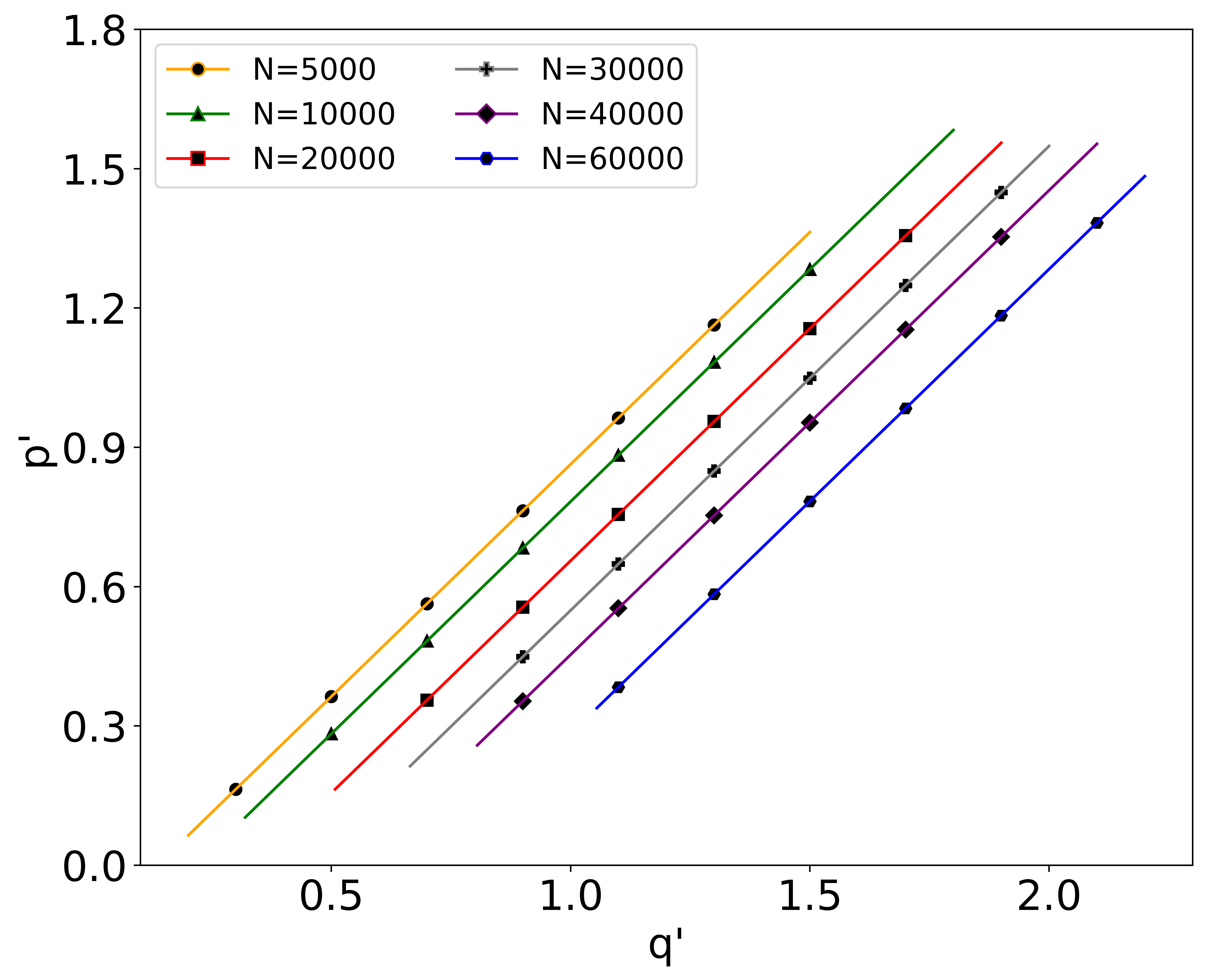}
    \label{fig:5e}
    }
    \subfloat[]{
    \includegraphics[width=0.32\linewidth]{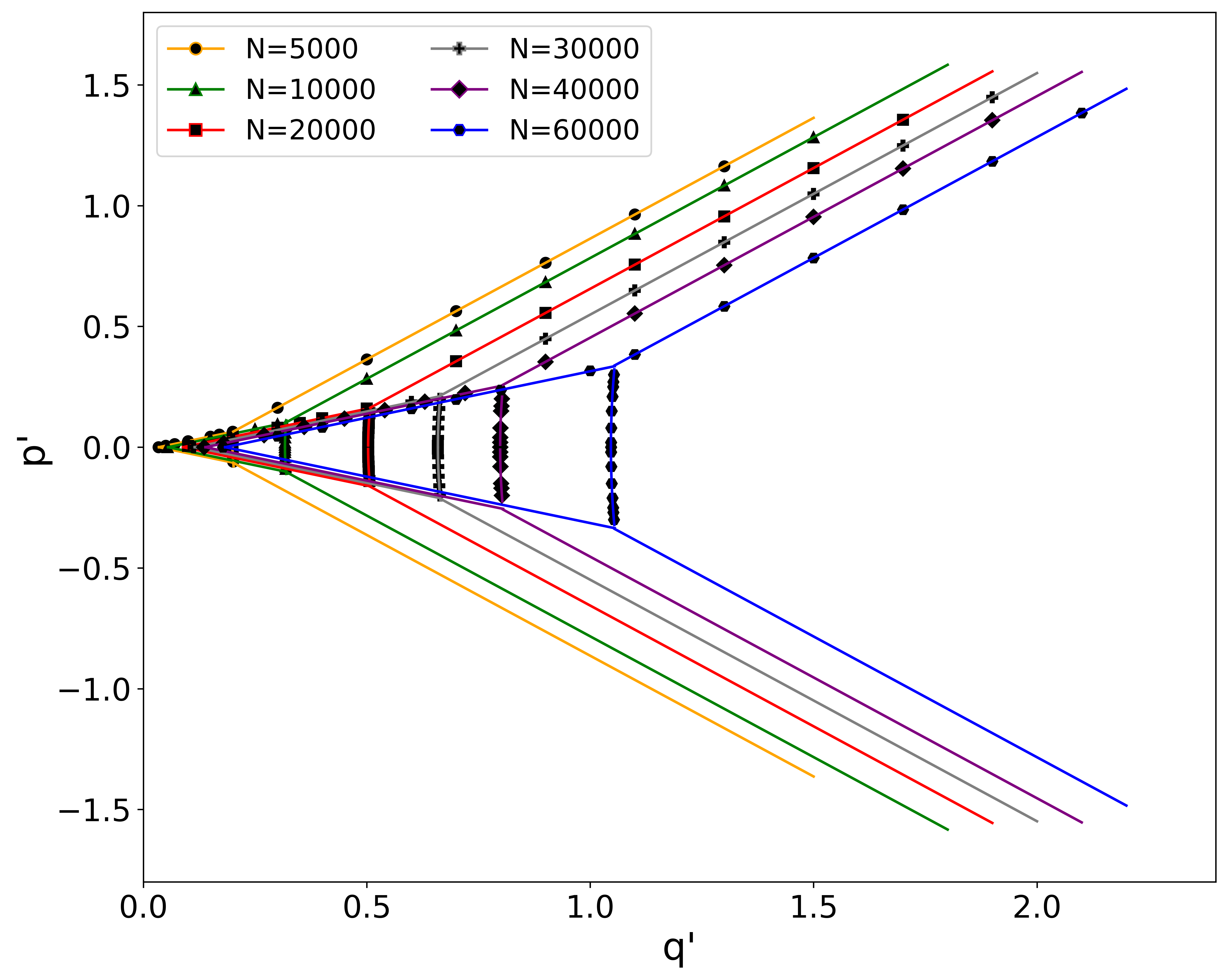}
    \label{fig:5f}
    }
    \\
    \subfloat[]{
    \includegraphics[width=0.32\linewidth]{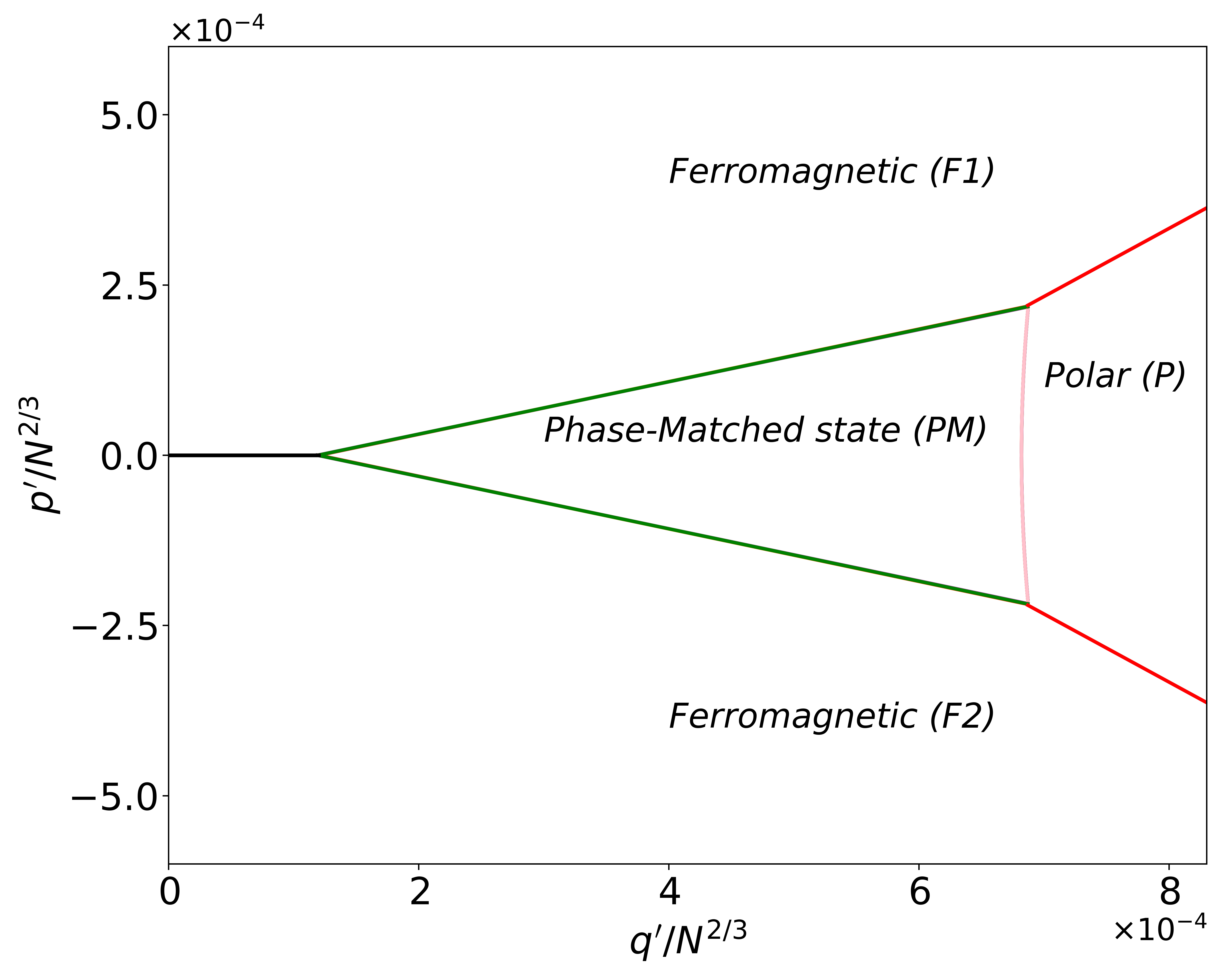}
    \label{fig:5g}
    }
    \caption{For the ferromagnetic type of spin interaction $\lambda_1<0$, (a) the phase boundary between the polar and PM phase for a trapped spin-1 condensate for different values of $N$ at $\lambda_1<0$. These phase boundaries are hyperbolic in nature and follow $p'^2=k q'^2-B(N,\lambda_1)$. (b) Variation of the scaling factor $B(N,\lambda_1)$ with $N$ for a fixed value of $\lambda_1$, shown in a log-log plot to get the power-law dependence. (c) Universal Polar-PM phase boundary obtained after scaling of $p'$ and $q'$ axes with $N^{2/3}$. The phase boundary between the (d) ferromagnetic and PM state, and (e) ferromagnetic and polar state, for different values of $N$. (f) The complete phase boundary for $\lambda_1<0$ for trapped condensates with varying values of $N$. (g) The universal phase diagram for $\lambda_1<0$, after scaling the $p'$ and $q'$ axes with $N^{2/3}$, shows the region of existence for different states.}
    \label{fig:5}
\end{figure*}

\subsubsection{PM-polar phase boundary}
We employ the VM to estimate the phase boundary between the PM and polar states. Note that this phase boundary is hyperbolic in the homogeneous case, as shown in Fig.\ref {fig:1}\subref{fig:1c}. We estimate the energies of the PM and polar states using VM for different system sizes (particle numbers). The resulting phase boundaries are shown in Fig.\ref{fig:5}\subref{fig:5a}. Although it may appear flat, in the trapped case the phase boundary also exhibits a hyperbolic shape, which is more evident in the inset plot. We use the hyperbola of type $p'^2=k q'^2-B(N,\lambda_1)$, where the constant turns out to be $k=6.48$. We then plot $B$ for different system sizes to confirm that $B(N,\lambda_1)$ has a power-law dependence on $N$ for a fixed $\lambda_1$. From the log-log plot in Fig.\ref{fig:5}\subref{fig:5b}, we get $B \propto N^{4/3}$ which results in the scaling factor $N^{2/3}$ for the $p^\prime$ and $q^\prime$.
With this scaling factor, the Polar-PM boundary collapse is shown in Fig.\ref{fig:5}\subref{fig:5c}.

\subsubsection{Ferromagnetic-PM phase boundary}
Following the same procedure, we find the phase boundary between the phase-matched state and the ferromagnetic state in Fig.\ref{fig:5}\subref{fig:5d}. 

These boundaries collapse under a linear fit, i.e., $p'=0.38q' - C$, where $C$ has an $N$ dependence that comes out to be $C\sim N^{2/3}$. Although the Ferromagnetic-PM phase boundary is a linear curve, it has a different functional form than the homogeneous predictions, where the Ferromagnetic-PM phase boundary is a $p=q$. An important observation is that the polar-PM and ferromagnetic-PM phase boundaries intersect at a particular value of $(q',p')$. Beyond this intersection point, we have to calculate the Polar-Ferromagnetic phase boundary, which is missing in the homogeneous results.

\subsubsection{Polar-Ferromagnetic phase boundary}
We can find the phase boundary for ferromagnetic and polar states for different values of $N$ (Fig.\ref{fig:5}\subref{fig:5e}). All these phase boundaries vary linearly, i.e, $p'=q'-D$, where $D\sim N^{2/3}$. 

We get to the complete phase boundary for different $N$ values (Fig.\ref{fig:5}\subref{fig:5f}).  Note that the F1-F2 phase boundary is a constant curve with $p'=0$. As we have already found, all individual phase boundaries exhibit universality when the linear and quadratic Zeeman terms are rescaled by $N^{2/3}$. Consequently, the universal phase diagram can be easily obtained using the same scaling (Fig.\ref {fig:5}\subref{fig:5g}).

\section{Discussion}
In this article, we present a new variational method for estimating the number density profiles of a spin-1 trapped condensate. The method presented here is easy to implement and helps to get the full phase diagram of the trapped condensate. For a quasi-one-dimensional harmonic confinement, we track all possible phase boundaries for varying numbers of condensate particles, considering both ferromagnetic and antiferromagnetic types of spin- interactions, thereby constructing the phase diagram. We find a suitable scaling that provides the universal feature of the phase diagram. We identify qualitative features of the trapped phase diagram that are completely absent in the homogeneous condensate. For the anti-ferromagnetic type of spin-interaction, the phase boundary between the anti-ferromagnetic and ferromagnetic states is independent of the quadratic Zeeman strength, according to the homogeneous result. We find that, for the trapped case, the same phase boundary becomes a function of the quadratic Zeeman term in the region where it is positive. On the other hand, for the ferromagnetic type of spin interaction, we find that the phase-matched state only exists for small values of both linear and quadratic Zeeman terms. This is in sharp contrast to the homogeneous condensate. This also results in a direct transition between the polar and ferromagnetic states for the trapped system. 

The method presented in this article is fairly general and can be easily extended to higher spin systems. The ansatz of continuous functions for the mean-fields (wavefunctions) makes this procedure fast and easy to implement. It is obvious that a better ansatz would make the estimated number density match the numerical results even more accurately, following the same procedure shown here. The simplifying assumptions we have taken to set some parameters in our ansatz of the wavefunction to be the same for all components can be relaxed to improve the accuracy. However, the computational cost for those little corrections could be substantial. It is important to note that probing near-boundary excitations does not require computations performed exactly at the transition region. Nevertheless, a reasonably accurate estimate of the transition region can help identify the parameter range in which the excitations responsible for the transition can be effectively probed. In this sense, the present variational method, with its easy implementation, can be quite effective in identifying regions of phase boundaries in trapped condensates. This establishes a unified first step towards future studies on instability and phase transitions near different phase boundaries and critical points of a trapped quasi-one-dimensional condensate. The method can also be extended to other types of harmonic confinement. 

\section{Acknowledgement}
The support and the resources provided by PARAM Brahma Facility under the National Supercomputing Mission, Government of India, at the Indian Institute of Science Education and Research, Pune, are gratefully acknowledged. PKK would like to acknowledge the support, in the form of a research fellowship, provided by the I-HUB Quantum Technology Foundation, Pune, India.

\bibliographystyle{apsrev4-1}
\bibliography{main} 	 
\appendix
\appendix

\section{Variational method for the Phase-matched state (PM state):}

For the PM state all three sub-components ($m=1,-1,0$) are populated with the condition $\theta_r=0$. For this multi-component state, we simplify our sub-component number density expression (Eq.\ref{eq10}) by assuming the parameters, $d_m$ and $b_m$, for all three sub-components are equal i.e., $d_1=d_{-1}=d_0=d$ and $b_1=b_{-1}=b_0=b$.
This leads to,
\begin{equation}
    \begin{split}
        u_m(\zeta)&=\left(a_m - b\zeta^2\right)^2\exp{\left(-\frac{\zeta^2}{d}\right)}\\
        &=(a_m^2+b^2\zeta^4-2a_mb\zeta^2)\exp{\left(-\frac{\zeta^2}{d}\right)},    
    \end{split}
\end{equation}
where $u_m(\zeta)$ is the sub-component densities. The total number density,
\begin{equation}\label{Eq:A2}
    \begin{split}
        u(\zeta)&=u_1(\zeta)+u_{-1}(\zeta)+u_0(\zeta)\\
        &=\Big((a_1^2+a_{-1}^2+a_0^2)+3b^2\zeta^4\\
        &\hspace{1cm}-2b(a_1+a_{-1}+a_0)\zeta^2\Big)\exp{\left(-\frac{\zeta^2}{d}\right)},
    \end{split}
\end{equation}
has now a total of six unknown parameters, namely $a_1,a_{-1},a_0,b,d$, and $\mu'$.

As discussed in detail in the context of AF state, for smaller values of $\zeta$, we match the total number density ansatz (Eq.\ref{Eq:A2}) with the corresponding TF expression for this state which leads to,
\begin{equation}
    (a_1^2+a_{-1}^2+a_0^2)=\frac{\mu'+ \frac{(p')^2-(q')^2}{2q'}}{\lambda_0+\lambda_1},
\end{equation}
\begin{equation}
    \mu'=\:(\lambda_0+\lambda_1)(a_1^2+a_{-1}^2+a_0^2)- \left(\frac{(p')^2-(q')^2}{2q'}\right),
\end{equation}
and
\begin{equation}
    b=\dfrac{\left(\frac{1}{2\lambda_1+2\lambda_0}-\frac{a_1^2}{d}-\frac{a_{-1}^2}{d}-\frac{a_0^2}{d}\right)}{2a_1+2a_0+2a_{-1}}.
\end{equation}

We are left with four parameters, namely $a_1$, $a_{-1}$, $a_0$, and $d$. We proceed with the same Lagrange multiplier method of free energy minimization,
\begin{equation}\label{eq:A7}
    N-\Bigg( \int n_{-1}(\zeta) \, d\zeta + \int n_1(\zeta) \, d\zeta  + \int n_0(\zeta) \, d\zeta\Bigg) =0,
\end{equation}
\begin{equation}\label{eq:A8}
\begin{split}
    \nabla_{a_m}\Bigg[E-\mu' &\Bigg( \int n_{-1}(\zeta) \, d\zeta\\ 
    &+ \int n_1(\zeta) \, d\zeta  + \int n_0(\zeta) \, d\zeta\Bigg)\Bigg]=0,
\end{split}
\end{equation}
The final step is to solve these four equations (Eq.\ref{eq:A7} and three equations for $m=1,0,-1$ in Eq.\ref{eq:A7}) simultaneously to get the remaining four parameters of the number density functions for a particular value of $q'$, $p'$, and $N$.
\end{document}